\documentclass[aps,prd,twocolumn,floatfix,superscriptaddress,nofootinbib]{revtex4-2}

\usepackage[english]{babel}
\usepackage[T1]{fontenc}
\usepackage{amsmath}
\usepackage{amssymb}
\usepackage{graphicx}
\usepackage[font=small,labelfont=bf,
   justification=justified,
   format=plain]{caption}
\usepackage{subcaption}
\usepackage{xcolor}
\usepackage[colorlinks]{hyperref}

\begin{document}

\title{Reconstructing isotropic and anisotropic \texorpdfstring{$f(\mathcal{Q})$}{} cosmologies}
\author{Fabrizio Esposito}
\email{fabrizio.esposito01@edu.unige.it}
\author{Sante Carloni}
\email{sante.carloni@unige.it}
\author{Roberto Cianci}
\email{roberto.cianci@unige.it}
\author{Stefano Vignolo}
\email{stefano.vignolo@unige.it}
\affiliation{DIME, Università di Genova, Via all'Opera Pia 15, 16145 Genova, Italy}

\begin{abstract}
We present a reconstruction algorithm for cosmological models based on $f(\mathcal{Q})$ gravity. We specifically focus on Bianchi Type-I and  Friedmann-Lema\^{\i}tre-Robertson-Walker spacetimes, obtaining exact solutions that might have application in a variety of scenarios such as spontaneous isotropization of Bianchi Type-I models, dark energy, and inflation as well as pre-Big Bang cosmologies.    
\end{abstract}
\date{\today}

\maketitle

\section{Introduction}\label{introduction}
General Relativity (GR) is a geometric theory of gravitation based on the Equivalence Principle \cite{Wald:1984rg, weinberg1972gravitation}. In GR, the dynamics of the gravitational field is described by the Einstein equations, which relate the curvature of spacetime to the energy-matter sources. 

Despite the great successes of Einstein's theory, some shortcomings of GR have emerged over the years. For example, in cosmology, no known matter source can generate the accelerated expansion phase that our Universe is experiencing. The currently most widely accepted model for cosmology, the $\Lambda$CDM model, assumes that the accelerated expansion is due to the cosmological constant $\Lambda$ \cite{Carroll_2001, Peebles_2003, Bamba:2012cp}. However, the observed value of $\Lambda$ disagrees with the theoretical prediction by 120 orders of magnitude \cite{RevModPhys.61.1}. This result led the community to consider more general fluids, generally known as dark energy, with the same key property as $\Lambda$, i.e. negative pressure. To this day, the nature of dark energy is still a matter of debate.  At the quantum level, the main conceptual problem is that in GR the metric is the field describing both the dynamics of gravity and the spacetime background. However, quantum theories are formulated on a fixed background. In order to solve the above problems, an interaction between geometry and quantum fields can be introduced, which is expressed by a modification of the Hilbert-Einstein Lagrangian through geometry-geometry or geometry-matter interaction terms. It is just in this field of research that the Extended Theories of Gravity were introduced \cite{CAPOZZIELLO2011167, Sotiriou_2010, Clifton_2012}.

In GR,  gravity is modeled in terms of geometrical properties of spacetime, represented by the Riemann tensor. The Riemann tensor (or more in general the curvature tensor), together with torsion and non-metricity, express the properties of a given affine connection defined on the spacetime. The set made up of curvature, torsion and non-metricity tensors is usually referred to as the ``Trinity of Gravity'' \cite{BeltranJimenez:2019tjy}. It has been known for some time that different representations of gravity can be obtained depending on the geometrical quantity considered, indeed, torsion and non-metricity contain enough degrees of freedom to describe the geometry of spacetime entirely. For instance, the Teleparallel Gravity (TG) \cite{Kr_k_2019} is a gravitational gauge theory where spacetime is assumed flat and torsion describes the gravitation through the Weitzenb\"{o}ck connection. Instead, the Symmetric Teleparallel Gravity (STG) \cite{Nester:1998mp,Adak:2008gd,Adak:2005cd, Bombacigno:2021bpk, Conroy:2017yln} describes gravity through a torsion-free and curvature-free connection but with non-metricity different from zero. As well as for GR, extensions of these theories have been developed, like the $f(T)$ (where $T$ is the torsion scalar) \cite{Cai:2015emx} and $f(\mathcal{Q})$ (where $\mathcal{Q}$ is the non-metricity scalar) theories  \cite{BeltranJimenez:2017tkd}.

A significant advantage of both $f(T)$ and  $f(\mathcal{Q})$ theories, with respect to other ``geometric'' extensions of GR (like $f(R)$ gravity), is that the background field equations are always of second order, so there are no instability problems related to the Ostrogradsky's theorem \cite{Motohashi_2015,Woodard:2015zca}.

In this paper, we focus on one of these recent modifications of GR: $f(\mathcal{Q})$ theory.  Several studies have been published on this kind of theory, mainly in connection with cosmological applications \cite{Harko_2018, Jim_nez_2020, Mandal_2020,Barros_2020,Lazkoz_2019, Soudi:2018dhv,Frusciante_2021,Hassan_2021,Bajardi_2020}. In particular, we aim to explore exact isotropic and anisotropic cosmological solutions using the so-called reconstruction methods.

Reconstruction methods were used for the first time by G. F. R. Ellis and M. S. Madsen \cite{Ellis_1991} to find the potential functions needed in models of inflationary universes. 
The general idea consists in reversing the usual procedure of resolution: a given form for the spatial scale factor is assumed and, once substituted into the cosmological equations, further information is derived on the remaining unknown functions of the theory, for instance, the inflaton potential function in \cite{Ellis_1991}.
Subsequently, reconstruction was used in various frameworks, like in the cosmology of $f(R)$ gravity \cite{Carloni_2012}, where the generic function of the Ricci scalar is reconstructed starting from given scale factors, in scalar-tensor cosmologies \cite{Vignolo_2013} and in the study of static and spherically symmetric spacetimes \cite{Carloni:2017rpu, Carloni:2017bck, Carloni:2014rba, 2021arXiv210205693N}.

In the following, we will consider Bianchi Type-I (BI) \cite{Ellis1969, Ellis2006} and spatially flat Friedmann-Lema\^{\i}tre-Robertson-Walker (FLRW) cosmologies with a focus on the role that non-metricity plays in cosmic history. The interest in the FLRW universe is evident due to its role in modern cosmology.  On the other hand, the BI metric, which corresponds to the simplest anisotropic generalization of the spatially flat FLRW spacetime, finds its interest in the analysis of the behavior of anisotropies in cosmology. In fact, even though the current Universe is believed to be essentially isotropic, it may not have been so at its beginning, nor does it necessarily have to be in the future. Furthermore, by studying less symmetrical metrics, we can better understand the isotropic ones, which can be considered their special subcase.

The layout of the paper is the following. In Sec. \ref{ch:f(Q)_theory} we review some generalities on $f(\mathcal{Q})$ theory. In Sec. \ref{ch:f(Q)_cosmology} we derive the cosmological equations for the BI metric, and then we restrict to the spatially flat FLRW one. In Sec. \ref{ch:recontruction_BI} and Sec.\ref{ch:recontruction_FRW} we apply the reconstruction method to $f(\mathcal{Q})$ cosmology. We devote Sec.  \ref{ch:conclusions} to a final discussion of the obtained results. Throughout the paper we use natural units ($c=8\pi G=1$) and the metric signature ($-,+,+,+$).

\section{\texorpdfstring{$f(\mathcal{Q})$ theory}{}} \label{ch:f(Q)_theory}
We assume a spacetime endowed with a metric tensor $g_{ij}$ and an affine connection $\Gamma_{ij}{}^{k}$ that introduces a corresponding covariant derivative $\nabla$. Given a metric $g_{ij}$, any affine connection can be decomposed as follows:
\begin{equation}
\Gamma_{ij}{}^{k} = \tilde{\Gamma}_{ij}{}^{k} + K_{ij}{}^{k} + N_{ij}{}^{k} ,
\label{eq:1_1}
\end{equation}
where $\tilde{\Gamma}_{ij}{}^{k}$ is the Levi-Civita connection,
\begin{equation}
\tilde{\Gamma}_{ij}{}^{k} = \frac{1}{2} g^{kh} \left( \partial_{i}g_{jh} + \partial_{j}g_{ih} - \partial_{k}g_{ij} \right),
\end{equation}
$K_{ij}{}^{k}$ is the contorsion tensor,
\begin{equation}
K_{ij}{}^{k} = \frac{1}{2} \left( T_{ij}{}^{k} - T_{i}{}^{k}{}_{j} - T_{j}{}^{k}{}_{i} \right),
\end{equation}
with the torsion tensor defined as
\begin{equation}
{T_{ij}}^{k} = {\Gamma_{ij}}^{k} - {\Gamma_{ji}}^{k},
\end{equation}
and $N_{ij}{}^{k}$ is the disformation tensor,
\begin{equation}
N_{ij}{}^{k} = \frac{1}{2} \left( Q^{k}{}_{ij} - Q_{i}{}^{k}{}_{j} - Q_{j}{}^{k}{}_{i} \right),
\end{equation}
defined in terms of the non-metricity tensor
\begin{equation}
Q_{kij} = \nabla_{k}g_{ij}.
\end{equation}
The curvature tensor associated with the connection $\Gamma_{ij}{}^{k}$ is expressed as
\begin{equation}
    R^{h}{}_{kij}= \partial_{i}\Gamma_{jk}{}^{h} - \partial_{j}\Gamma_{ik}{}^{h} + \Gamma_{ip}{}^{h}\Gamma_{jk}{}^{p} - \Gamma_{jp}{}^{h}\Gamma_{ik}{}^{p}. 
\end{equation}
We will focus on $f(\mathcal{Q})$ gravity, namely a generalization of STG based on an action of the form
\begin{equation}
A = \int d^4x \left[-\frac{1}{2} \sqrt{-g} f(\mathcal{Q}) + \lambda_{a}{}^{bij}R^{a}{}_{bij} + \lambda_{a}{}^{ij}T_{ij}{}^{a} \right] + A_{m}
\end{equation}
where $A_{m}$ indicates a generic matter action, $\lambda_{a}{}^{bij}$ and $\lambda_{a}{}^{ij}$ are Lagrange multipliers introduced to impose the vanishing of curvature and torsion, and $f(\mathcal{Q})$ is a generic function of the non-metricity scalar,
\begin{equation}\label{scalarQ}
\begin{aligned}
\mathcal{Q} =& -Q_{hij}P^{hij} =\\
=& \frac{1}{4}Q_{hij}Q^{hij} - \frac{1}{2}Q_{hij}Q^{ijh} - \frac{1}{4} q_{h}q^{h} + \frac{1}{2}q_{h}Q^{h},
\end{aligned}
\end{equation}
In Eq. \eqref{scalarQ}, the tensor
\begin{equation}
\begin{aligned}
P^{h}{}_{ij} =& - \frac{1}{4} Q^{h}{}_{ij} + \frac{1}{2} Q_{(ij)}{}^{h} + \frac{1}{4} q^{h} g_{ij} - \frac{1}{4} Q^{h} g_{ij} - \frac{1}{4} \delta^{h}_{(i} q_{j)}
\end{aligned}
\end{equation}
is defined as the conjugate of the non-metricity tensor, and
\begin{equation}
    q_{h} = Q_{hi}{}^{i} \quad \mbox{and} \quad Q_{h}=Q_{ih}{}^{i}.
\end{equation}
are the two independent traces of $Q_{hij}$.
The expression for the non-metricity scalar \eqref{scalarQ} has been chosen in such a way that when setting $f(\mathcal{Q})=\mathcal{Q}$ we obtain a theory that is equal to GR  modulo a boundary term. 

We develop the theory in the metric-affine framework where metric and connection are independent variables. By varying with respect to the Lagrange multipliers, we obtain the constraints
\begin{equation}\label{curvature_torsion_free}
R^{a}{}_{bij} = 0 \qquad \mbox{and} \qquad T_{ij}{}^{a} = 0,
\end{equation}
whereas variations with respect to the metric and the connection yield field equations of the form:
\begin{equation} \label{eq:metric_equation}
\begin{split}
\frac{2}{\sqrt{-g}}\nabla_{h} \left( \sqrt{-g} f' P^{h}{}_{ij} \right) + \frac{1}{2}g_{ij}f(\mathcal{Q})  +\quad\\ 
+ f' \left( P_{iab}Q_{j}{}^{ab} - 2 Q^{ab}{}_{i}P_{abj} \right) = \Sigma_{ij}
\end{split}
\end{equation}
and
\begin{equation}\label{eq:connection_equation}
 \nabla_{p} \lambda_{h}{}^{jip} +  \lambda_{h}{}^{ij} - \sqrt{-g} f' P^{ij}{}_{h} = \Phi^{ij}{}_{h},
\end{equation}
with 
\begin{equation}
    \Sigma_{ij}=  -\frac{2}{\sqrt{-g}}\frac{\delta  \mathcal{L}_{m}}{\delta g^{ij}} \quad \text{and} \quad
    \Phi^{ij}{}_{h} = - \frac{1}{2}\frac{\delta \mathcal{L}_{m}}{\delta {\Gamma_{ij}{}^{h}}}.
\end{equation}
$\mathcal{L}_{m}$ is a generic matter Lagrangian density that includes $\sqrt{-g}$ in its definition.

The flatness and torsionless conditions \eqref{curvature_torsion_free} ensure the existence of local coordinates in which $\Gamma_{ij}{}^{h}=0$. In the following, we will systematically adopt this choice of coordinates, usually referred to as the ``coincident gauge.'' We also note that, due to the flatness and torsionless conditions, Eq. \eqref{eq:connection_equation} can be written as follows:
\begin{equation}\label{eq:connection_equation_2}
    \nabla_{i}\nabla_{j}\left(\sqrt{-g} f' P^{ij}{}_{h}\right) +\nabla_{i}\nabla_{j} \Phi^{ij}{}_{h}=0 .
\end{equation}
Using Eq. \eqref{eq:connection_equation_2} into the Levi-Civita divergence of Eq. \eqref{eq:metric_equation}, we derive the energy-momentum conservation:
\begin{equation}\label{energy-momentum_conservation}
\tilde{\nabla}_{i}\Sigma^{i}{}_{h} + \frac{2}{\sqrt{-g}}\nabla_{i}\nabla_{j} \Phi^{ij}{}_{h} = 0,
\end{equation}
where $\tilde{\nabla}$ denotes the Levi-Civita covariant derivative. Since we will consider only perfect fluids as source of the gravitational field equations, we will assume that the matter action does not depend on the connection. Therefore, from now on we will consider $\Phi^{ij}{}_{h}=0$. 

\section{\texorpdfstring{$f(\mathcal{Q})$ cosmology} {}}\label{ch:f(Q)_cosmology}
Let us now specialize the field equations of \eqref{eq:metric_equation} to the study of cosmological models. In particular, we focus on BI and FLRW universes. These equations will constitute the base of the reconstruction method we intend to apply. 

\subsection{Bianchi Type-I metric}
The BI metric represents spatially flat, homogeneous but not isotropic spacetimes. The most common realization of this metric is
\begin{equation}\label{eq:bianchi_metric}
    ds^2 = - dt^2 + a^2(t)dx^2 + b^2(t)dy^2 + c^2(t)dz^2,
\end{equation}
where $a(t)$, $b(t)$ and $c(t)$ are the scale factors associated to each space direction.
Remembering we are using the coincident gauge, in this metric the non-metricity scalar is
\begin{equation}\label{eq:bianchi_nonmetricity}
    \mathcal{Q} =  2 \left( \frac{\dot{a}\dot{b}}{ab} + \frac{\dot{a}\dot{c}}{ac} +\frac{\dot{b}\dot{c}}{bc} \right).
\end{equation}

\subsubsection{Cosmological equations}
To derive the cosmological equations, we assume that, at the cosmological level, matter is described by the energy-momentum tensor of a perfect fluid,
\begin{equation}
    \Sigma_{ij} = \left( \rho + p \right) u_{i}u_{j} + p g_{ij},
\end{equation}
where pressure $p$ and the energy density $\rho$ are related by the equation of state $p=w\rho$, with $w$ the barotropic factor, and $\rho$ satisfies the continuity equation derived from Eq. \eqref{energy-momentum_conservation},
\begin{equation} \label{eq:continuity_BI}
    \dot{\rho} + \frac{\dot{\tau}}{\tau} \left( 1 + w \right)\rho = 0,
\end{equation}
whose solution in terms of the volume of the Universe $\tau(t)=a(t)b(t)c(t)$ is
\begin{equation}
    \rho = \rho_{0} \tau^{-\left(1+w\right)},
\end{equation}
where $\rho_{0}$ is the density at a given initial time. We derive the cosmological equations from the temporal and spatial part of Eq. \eqref{eq:metric_equation}, whereas Eq. \eqref{eq:connection_equation_2} is identically satisfied,
\begin{equation} \label{eq:BI_0}
    \frac{1}{2} f - 2f' \left( \frac{\dot{a}\dot{b}}{ab} + \frac{\dot{b}\dot{c}}{bc} + \frac{\dot{a}\dot{c}}{ac} \right) = -\rho ,
\end{equation}
\begin{equation}\label{eq:BI_1}
    \dot{f}' \left( \frac{\dot{a}}{a} - \frac{\dot{\tau}}{\tau} \right) - f' \left( \frac{\ddot{b}}{b} + \frac{\ddot{c}}{c} + \frac{\dot{a}\dot{b}}{ab} + \frac{\dot{a}\dot{c}}{ac} + 2 \frac{\dot{b}\dot{c}}{bc} \right) + \frac{1}{2}f = p,
\end{equation}
\begin{equation}\label{eq:BI_2}
    \dot{f}' \left( \frac{\dot{b}}{b} - \frac{\dot{\tau}}{\tau} \right) - f' \left( \frac{\ddot{a}}{a} + \frac{\ddot{c}}{c} + \frac{\dot{a}\dot{b}}{ab} + \frac{\dot{b}\dot{c}}{bc} + 2 \frac{\dot{a}\dot{c}}{ac} \right) + \frac{1}{2}f = p,
\end{equation}
\begin{equation}\label{eq:BI_3}
    \dot{f}' \left( \frac{\dot{c}}{c} - \frac{\dot{\tau}}{\tau} \right) - f' \left( \frac{\ddot{a}}{a} + \frac{\ddot{b}}{b} + \frac{\dot{a}\dot{c}}{ac} + \frac{\dot{b}\dot{c}}{bc} + 2 \frac{\dot{a}\dot{b}}{ab} \right) + \frac{1}{2}f = p.
\end{equation}
We can recast the above equations in a more useful form by performing some simple operations.

Let us consider the combination of the Eqs. \eqref{eq:BI_0}-\eqref{eq:BI_3} given by
Eq. \eqref{eq:BI_0} multiplied by $-\dot{\tau}/\tau$ and added to Eqs. \eqref{eq:BI_1}, \eqref{eq:BI_2}, and \eqref{eq:BI_3} multiplied by $3\dot{a}/a$, $3\dot{b}/b$, and $3\dot{c}/c$, respectively.
The resulting expression is an equivalent of the Raychaudhuri equation,
\begin{equation}\label{eq:raychaudhuri_BI}
    f\frac{\dot{\tau}}{\tau} - f'\left( \frac{3}{2}\dot{\mathcal{Q}} + 2 \mathcal{Q}\frac{\dot{\tau}}{\tau} \right) - 3 f''\mathcal{Q} \dot{\mathcal{Q}} = \frac{\dot{\tau}}{\tau}\left( \rho + 3p \right).
\end{equation}
However, we can also obtain Eq. \eqref{eq:raychaudhuri_BI} using only Eqs. \eqref{eq:continuity_BI} and \eqref{eq:BI_0} taking the time derivative of Eq. \eqref{eq:BI_0} multiplied by $3$ and adding Eq. \eqref{eq:BI_0} multiplied by $2\dot{\tau}/\tau$.
Thus, if the Eqs. \eqref{eq:continuity_BI} and \eqref{eq:BI_0} are satisfied, so is Eq. \eqref{eq:raychaudhuri_BI}. This step will be crucial for developing the reconstruction algorithm as it allows us to remove one equation.

Instead, subtracting Eq. \eqref{eq:BI_2} and Eq. \eqref{eq:BI_3} from Eq. \eqref{eq:BI_1}, we get, respectively:
\begin{equation}
    \frac{\dot{a}}{a} - \frac{\dot{b}}{b} = \frac{k_{ab}}{f' \tau} 
    \quad \to \quad \frac{a}{b} = e^{d_{1}}\exp \int \frac{k_{ab}}{f' \tau} dt
\end{equation}
and
\begin{equation}
    \frac{\dot{a}}{a} - \frac{\dot{c}}{c} = \frac{k_{ac}}{f' \tau} 
    \quad \to \quad \frac{a}{c} = e^{d_{2}}\exp \int \frac{k_{ac}}{f' \tau} dt,
\end{equation}
where $k_{ab}, k_{ac}, d_1$, and $d_2$ are constants of integration.

We can, therefore, consider the equivalent set of independent cosmological equations given by:
\begin{equation} \label{eq:cosmology_2}
    \dot{\rho} + \frac{\dot{\tau}}{\tau} \left( 1 + w \right)\rho = 0,
\end{equation}
\begin{equation}\label{eq:cosmology_1}
    \frac{1}{2} f - 2f' \left( \frac{\dot{a}\dot{b}}{ab} + \frac{\dot{b}\dot{c}}{bc} + \frac{\dot{a}\dot{c}}{ac} \right) = -\rho ,
\end{equation}
\begin{equation}\label{eq:cosmology_4}
    \frac{\dot{a}}{a} - \frac{\dot{b}}{b} = \frac{k_{ab}}{f' \tau},
\end{equation}
\begin{equation}\label{eq:cosmology_5}
    \frac{\dot{a}}{a} - \frac{\dot{c}}{c} = \frac{k_{ac}}{f' \tau}.
\end{equation}
One of the last two equations can be replaced, depending on the situation we are analyzing, by the relation that we derive with simple algebraic steps from Eqs. \eqref{eq:cosmology_4} and  \eqref{eq:cosmology_5},
\begin{equation}\label{eq:cosmology_6}
(k_{ab}-k_{ac})\frac{ \dot{a}}{a}+k_{ac}\frac{ \dot{b}}{b}-k_{ab}\frac{\dot{c}}{c} = 0.
\end{equation}
Notice that the number of  cosmological equations has decreased. This is possible because one of the Eqs. \eqref{eq:BI_1}-\eqref{eq:BI_3} can be replaced by Eq. \eqref{eq:raychaudhuri_BI}, which is always satisfied given the solutions of the Eqs. \eqref{eq:continuity_BI} and \eqref{eq:BI_0}.

\subsection{FLRW metric}
If we require  isotropy in the BI metric, that is $a(t)=b(t)=c(t)$ in \eqref{eq:bianchi_metric}, we obtain the spatially flat FLRW metric. The non-metricity scalar is related to the Hubble parameter $H = \dot{a}/{a}$ by
\begin{equation}
  \mathcal{Q} = 6 H^{2}
\end{equation}
and the Raychaudhuri equation \eqref{eq:raychaudhuri_BI} is equal to
\begin{equation}\label{eq:raychaudhuri_FLRW}
    \frac{1}{6}f - f'\left( \dot{H} + 2 H^{2} \right) - 12 f'' H^{2} \dot{H} = \frac{1}{6}\left(\rho + 3p\right).
\end{equation}
The cosmological equations \eqref{eq:cosmology_2}-\eqref{eq:cosmology_5} reduce to the set
\begin{equation}\label{eq:cosmology_FLRW_1}
\frac{1}{2}f-6H^{2}f'=-\rho,
\end{equation}
\begin{equation}\label{eq:cosmology_FLRW_2}
\dot{\rho} + 3H(1+w)\rho=0.
\end{equation}
In order to facilitate the study of the examples that will be considered, it is useful to define the deceleration parameter,
\begin{equation}
    q = -\frac{\ddot{a}a}{\dot{a}^{2}},
\end{equation}
and rewrite Eqs. \eqref{eq:raychaudhuri_FLRW} and \eqref{eq:cosmology_FLRW_1} in a more expressive form,
\begin{gather}
     \frac{\ddot{a}}{a} = -\frac{1}{6}\left(\hat{\rho}_{M} + 3\hat{p}_{M} \right) - \frac{1}{6}\left(\hat{\rho}_{f} + 3\hat{p}_{f}\right), \label{eq:raychaudhuri_FLRW_effective}\\
    H^{2} = \frac{1}{3} \left(\hat{\rho}_{M} + \hat{\rho}_{f}\right), \label{eq:cosmology_FLRW_1_effective}
\end{gather}
where 
\begin{equation}
    \hat{\rho}_{M} = \frac{1}{2}\frac{\rho}{f'} \qquad \text{and} \qquad \hat{p}_{M} = \frac{p}{f'}
\end{equation}
represent the standard energy density and pressure, with an effective gravitational constant regulated by $f'(\mathcal{Q})$, while 
\begin{equation}
    \hat{\rho}_{f} = \frac{1}{4} \frac{f}{f'}, \quad \text{and} \quad
    \hat{p}_{f}= 2\left[ \frac{\mathcal{Q}}{4}-\frac{1}{4}\frac{f}{f'} +\frac{f''}{f'}H\dot{\mathcal{Q}} \right]
\end{equation}
represent the energy density and pressure of an effective fluid associated with the presence of non-metricity.

\section{Reconstruction method: Bianchi Type-I}\label{ch:recontruction_BI}
In this section, we will apply the reconstruction algorithm to investigate some exact cosmological models of the type BI. Given suitable scale factors, we will find the function $f(\mathcal{Q})$, which admits such scale factors as solutions of the corresponding cosmological equations.  

\subsection{Example 1: Power law scale factors}\label{subsec:BI_Example_1}
Let us start by assuming each scale factor as a power law as in the classical Kasner solution \cite{stephani2004relativity} (but without restrictions on the exponents),
\begin{gather}
    a(t)=a_{0}t^{n} \qquad b(t)=b_{0}t^{m}, \qquad c(t)=c_{0}t^{l} \label{eq:exampleBI_1_1a}, \\ 
    \tau = abc = a_{0}b_{0}c_{0}t^{N} \label{eq:exampleBI_1_1b}
\end{gather}
where $N=n+m+l$ and $a_{0}$, $b_{0}$ and $c_{0}$ are dimensional constants. In such a circumstance, the non-metricity scalar takes the form
\begin{equation}\label{IV_Q}
    \mathcal{Q}=2\left(nm+nl+ml\right)t^{-2} = \xi t^{-2},
\end{equation}
with $\xi=2(nm+nl+ml)$. Making use of Eq. \eqref{IV_Q}, from the definition of the spatial volume and the continuity equation we obtain the expressions of  $\tau$ and $\rho$ as functions of $\mathcal{Q}$,
\begin{equation}
    \tau(\mathcal{Q}) = \tau_{0} \left(\frac{\xi}{\mathcal{Q}}\right)^{\frac{N}{2}},
\end{equation}
\begin{equation}\label{IV_density}
\rho(\mathcal{Q}) =\rho_{0}\tau^{-(1+\bar{w})} = \rho_{0} \tau_{0}^{-(1+\bar{w})}\left( \frac{\mathcal{Q}}{\xi}\right)^{\frac{1}{2}(1+\bar{w})N},
\end{equation}
where $\tau_{0}=a_{0}b_{0}c_{0}$. 
In the above expression, and in the following, we will use $\bar{w}$ instead of $w$ to emphasize the fact that $\bar{w}$ is just a parameter for the theory we will reconstruct, and it is not related to any matter source the final reconstructed theory might be coupled with. 
Inserting Eq. \eqref{IV_density} into Eq. \eqref{eq:cosmology_1}, we get the differential equation
\begin{equation}\label{IV_f(Q)_1}
\frac{f(\mathcal{Q})}{2}-\mathcal{Q} f'(\mathcal{Q})= 
- \epsilon \mathcal{Q}^{\frac{1}{2} (1+\bar{w}) N}
\end{equation}
with
\begin{equation}
\epsilon = \rho_{0} \tau_{0}^{-(1+\bar{w})} \xi^{-\frac{1}{2} (1+\bar{w}) N}.
\end{equation}
Equation \eqref{IV_f(Q)_1} admits the solution,
\begin{equation}\label{IV_solutionf1}
f(\mathcal{Q}) = f_{0} \sqrt{\mathcal{Q}} + 2 \epsilon\frac{ \mathcal{Q}^{\frac{1}{2} (1+\bar{w}) N}}{N(1+\bar{w})-1},
\end{equation}
where $f_{0}$ is a constant of integration and will be so throughout the paper\footnote{One might think that the result in Eq. \eqref{IV_solutionf1} is only valid for the fluid chosen in the reconstruction process, however, such a conclusion would be incorrect. In fact, if we use Eq. \eqref{IV_solutionf1} and fluids with $w \neq \bar{w}$ in Eqs. \eqref{eq:BI_0}-\eqref{eq:BI_3}, then we obtain a different evolution for the scale factors from the one used for the reconstruction method. For the sake of simplicity, we will show this explicitly in Section \ref{ch:recontruction_FRW} for the FRLW case.}.

Using Eqs. \eqref{eq:exampleBI_1_1a}, \eqref{eq:exampleBI_1_1b} and \eqref{IV_solutionf1}, Eqs. \eqref{eq:cosmology_4}, and \eqref{eq:cosmology_5} generate the following constraints on the integration constants:
\begin{equation}\label{eq:exampleBI_1_con1}
   f_{0}=0, \qquad n + m + l = \frac{1}{\bar{w}},
\end{equation}
\begin{equation}\label{eq:exampleBI_1_con2}
    k_{ab} = \frac{\bar{w} (1+\bar{w}) \rho_{0} \tau_{0}^{-\bar{w} } (n-m)}{2 \left[m+n-\bar{w}  \left(m^2+m n+n^2\right)\right]},
\end{equation}
\begin{equation}\label{eq:exampleBI_1_con3}
    k_{ac} = \frac{(1+\bar{w}) \rho_{0} \tau_{0}^{-\bar{w} } (\bar{w}  m+2 \bar{w}  n-1)}{2 \left[m+n-\bar{w}  \left(m^2+m n+n^2\right)\right]}.
\end{equation}
Notice that relation \eqref{eq:exampleBI_1_con1} implies that we are forced to exclude the case $\bar{w}=0$. If we set $\bar{w}=0$ from the beginning, then we would obtain:
\begin{equation}
    m=n=l=\frac{1}{3},
\end{equation}
i.e. an isotropic solution. 

\subsection{Example 2: More complex scale factors.}\label{subsec:BI_Example_2}
In this second example, we choose the scale factors as follows:
\begin{equation}
    \begin{split}
        a(t) = \alpha \texttt{a}(t) \sqrt[3]{\tau (t)},  \qquad b(t) = \beta \texttt{b}(t) \sqrt[3]{\tau (t)}, \\
        c(t) = \gamma \texttt{c}(t) \sqrt[3]{\tau (t)}, \qquad \qquad \ \ \
    \end{split}
\end{equation}
which is an interesting template for BI solutions (see, e.g. \cite{stephani2004relativity}). The quantities $\alpha$, $\beta$, and $\gamma$ are generic constants.

The first step is to derive $\texttt{b}(t)$ from Eq. \eqref{eq:cosmology_6},
\begin{equation}
   \texttt{b} = b_{0} \texttt{a}^{\frac{k_{ac}-k_{ab}}{k_{ac}}} \texttt{c}^{\frac{k_{ab}}{k_{ac}}},
\end{equation}
where $b_{0}$ is a constant of integration. Then, using the definition of $\tau$ and $\mathcal{Q}$, we obtain the scale factor
\begin{equation}
    \texttt{a} =  a_{0}\texttt{c}^{\frac{k_{ab}+k_{ac}}{k_{ab}-2k_{ac}}}
\end{equation}
and the relation
\begin{equation}
    \frac{\dot{\texttt{c}}^{2}}{\texttt{c}^{2}} = \frac{(k_{ab}-2 k_{ac})^{2} }{18\Omega^{2}}\left(2 \frac{\dot{\tau}^2}{\tau^{2}} - 3 \mathcal{Q}\right)
\end{equation}
with
\begin{equation}
    a_{0} = \left( b_{0} \alpha \beta \gamma \right)^{\frac{k_{ac}}{k_{ab}-2k_{ac}}}  
\end{equation}
and 
\begin{equation}
    \Omega = \sqrt{k_{ab}^2-k_{ab} k_{ac}+k_{ac}^2}.
\end{equation}
If we now extrapolate $f'(\mathcal{Q})$ from Eq. \eqref{eq:cosmology_1},
\begin{equation}
    f' = \frac{f+2 \rho_{0} \tau^{-(1+\bar{w})}}{2 \mathcal{Q}}, 
\end{equation}
and replace all in Eq. \eqref{eq:cosmology_4}, then we find that particular cosmological solutions 
can be found imposing the conditions,
\begin{equation}
    f = K_{f} \tau^{-1-\bar{w}}, \label{eq:exampleBI_2_1}
\end{equation}
\begin{equation}
\frac{\dot{\tau}^2}{\tau^2}=\frac{1}{4} \mathcal{Q} \left(K_{\tau}^2 \ \mathcal{Q} \ \tau^{2 \bar{w}}+6\right), \label{eq:exampleBI_2_2}
\end{equation}
\begin{equation}
    K_{\tau} (K_{f}+2 \rho_{0})-4 \Omega=0. \label{eq:exampleBI_2_3}
\end{equation}
Equations \eqref{eq:exampleBI_2_1} and \eqref{eq:exampleBI_2_2} can be resolved if we make explicit the dependence of $\mathcal{Q}$ on $\tau$. A convenient choice is 
\begin{equation}\label{IV_tau}
    \tau = \pm \left(\frac{\mathcal{Q}_{0}}{\mathcal{Q}}\right)^{\frac{1}{n}},
\end{equation}
\begin{equation}
    f(\mathcal{Q}) = K_{f} \left[\pm\left(\frac{\mathcal{Q}_{0}}{\mathcal{Q}}\right)^{\frac{1}{n}}\right]^{-1-\bar{w}},
\end{equation}
which, when substituted into Eq. \eqref{eq:cosmology_1}, provides the value of the constant $K_{f}$,
\begin{equation}
    K_{f} = \frac{2 n \rho_{0}}{2 \bar{w} + 2 - n}
\end{equation}
and, from Eq.\eqref{eq:exampleBI_2_3}, also the value of $K_{\tau}$,
\begin{equation}
    K_{\tau} = \frac{\Omega (2\bar{w} + 2 -n)}{\rho_{0} (1+\bar{w})}.
\end{equation}
To proceed further, we will analyze separately three subcases: $\bar{w}=0$, $n=2\bar{w}$, and $n=\bar{w}$.

\subsubsection{Case: \texorpdfstring{$\bar{w}=0$}{}}\label{subsec:BI_Example_2_1}
\begin{figure}[ht]
    \begin{subfigure}[ht]{\linewidth}
        \includegraphics[width=8.6cm, height=4.5cm]{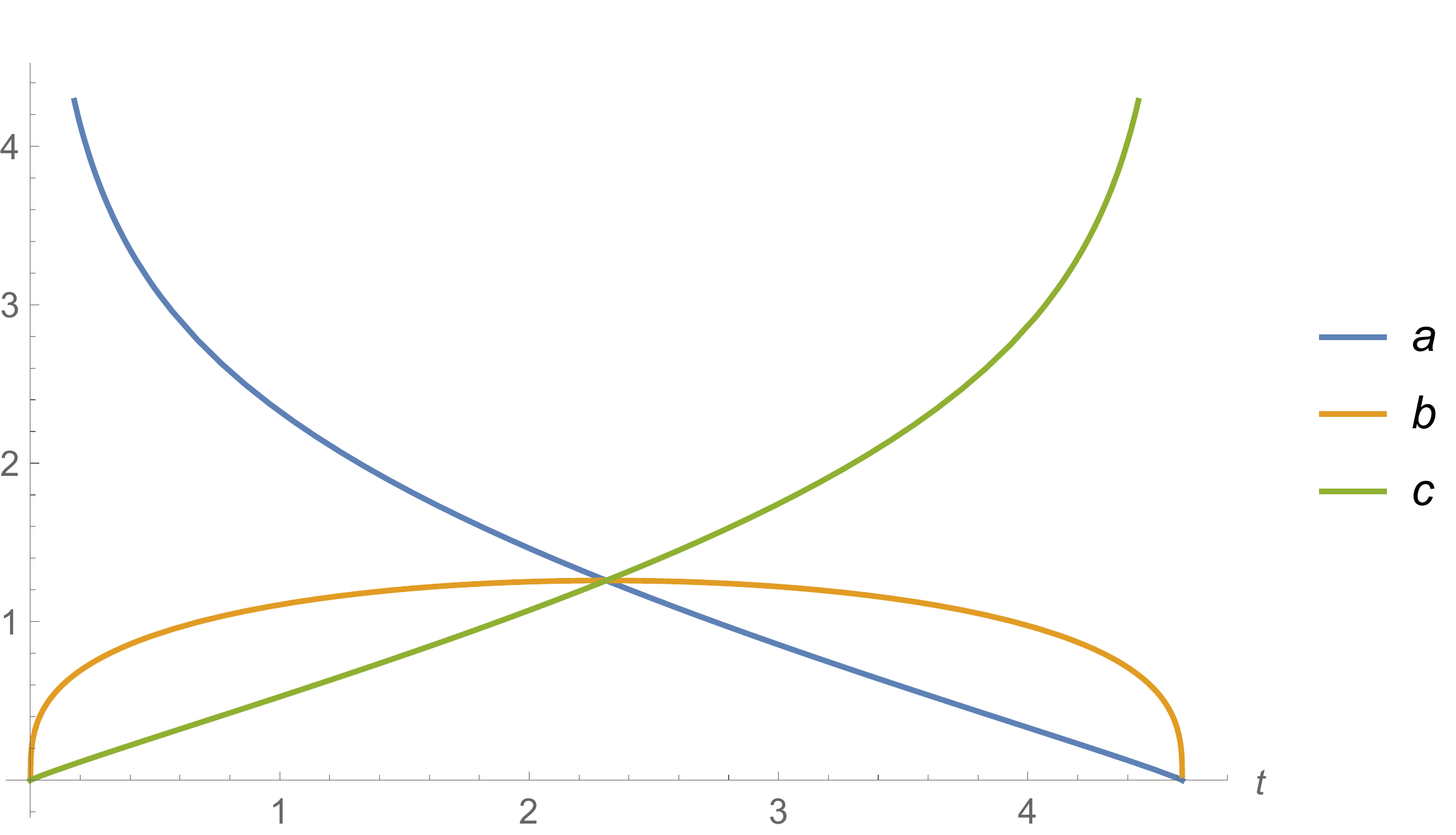}
        \caption{Scale factors}
        \label{fig:BI_example2_1_a}
    \end{subfigure}
    \begin{subfigure}[ht]{\linewidth}
        \includegraphics[width=8.6cm, height=4.5cm]{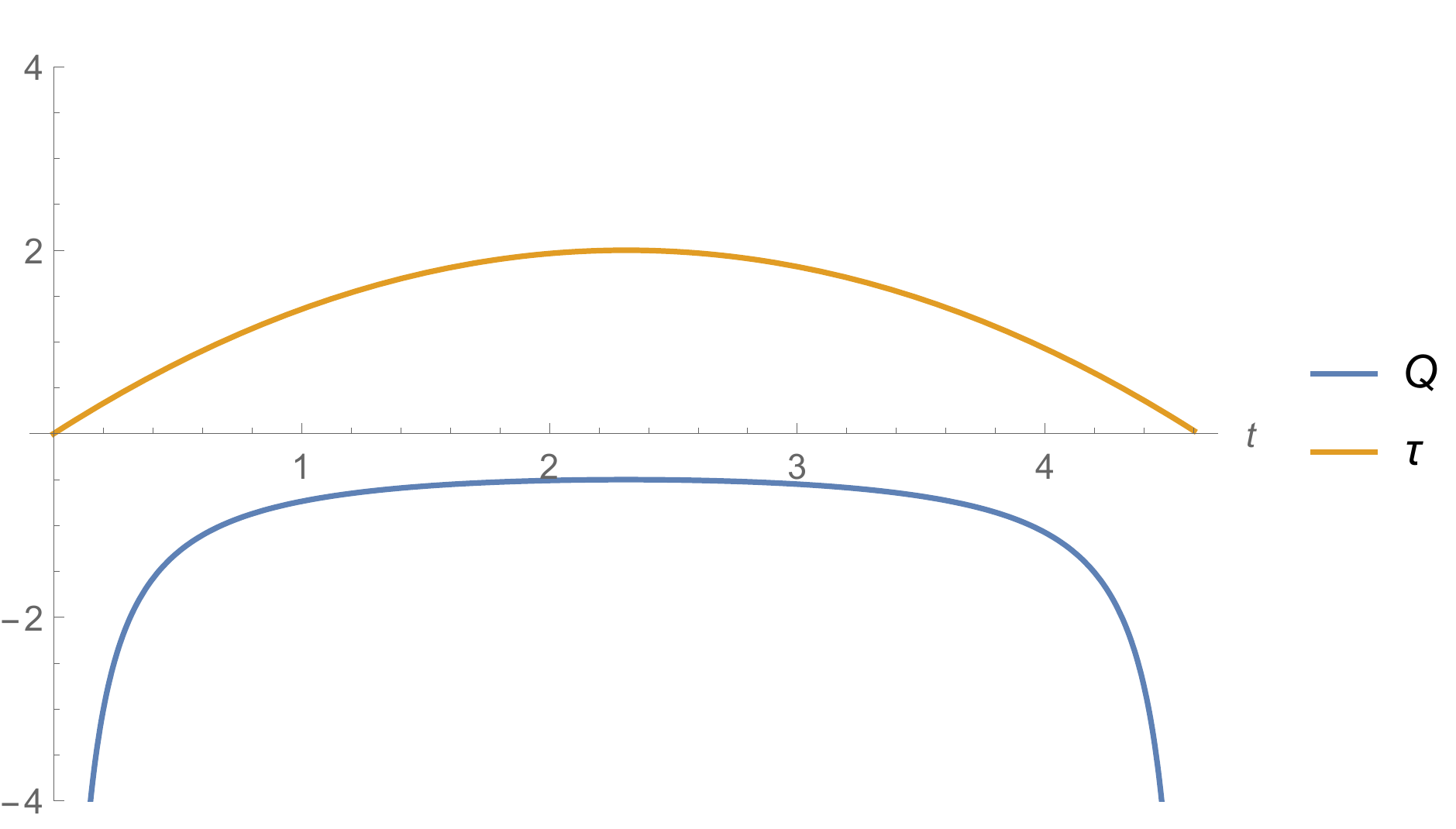} 
        \caption{Non-metricity scalar and $\tau$.}
        \label{fig:BI_example2_1_b}
    \end{subfigure}
\caption{Evolution of \eqref{eq:tau_BI_2_1}-\eqref{eq:c_BI_2_1} with  values $n=1$, $K_{\tau}=2\sqrt{3}$, $\mathcal{Q}_{0}=1$, $a_{1}=b_{1}=c_{1}=\frac{\sqrt[3]{3}}{2}$, $m_{1}=\frac{2}{\sqrt{3}}$, $m_{2}=0$, $m_{3}=-\frac{2}{\sqrt{3}}$, and $t_{0}=\frac{4}{\sqrt{3}}$.}
    \label{fig:BI_example2_1}
\end{figure}

If we set $\bar{w}=0$ and
\begin{equation}
    \tau = - \left(\frac{\mathcal{Q}_{0}}{\mathcal{Q}}\right)^{\frac{1}{n}},
\end{equation}
with $n$ an odd integer, then the solution of Eq. \eqref{eq:exampleBI_2_2} is
\begin{equation}\label{eq:tau_BI_2_1}
    \tau(t) = a_{1}b_{1}c_{1} \left[ \frac{4K_{\tau}^2}{9n^{2}} -(t-t_{0})^2\right]^{\frac{1}{n}}.
\end{equation}
The quantities $a_{1}$, $b_{1}$ and $c_{1}$ are constants depending on $\alpha$, $\beta$, $\gamma$, $b_{0}$, $k_{ab}$, $k_{ac}$, $\rho_{0}$ and $\mathcal{Q}_{0}$ and the parameter $t_{0}$ is the instant of time in which the initial data are assigned. This notation will also be used in the subsequent examples.
The scale factors assume the form,
\begin{equation}\label{eq:a_BI_2_1}
\begin{aligned}
    a(t) =& a_{1} \left[\frac{4K_{\tau}^2}{9n^{2}} -(t-t_{0})^2\right]^{\frac{1}{3 n}} \\
    &\exp \Bigg\{-m_{1}\tanh ^{-1}\left[\frac{3 n (t-t_{0})}{2 K_{\tau}}\right]\Bigg\},
\end{aligned}
\end{equation}
\begin{equation}\label{eq:b_BI_2_1}
\begin{aligned}
    b(t) =& b_{1} \left[\frac{4K_{\tau}^2}{9n^{2}} -(t-t_{0})^2\right]^{\frac{1}{3 n}}\\
    &\exp \Bigg\{-m_{2} \tanh ^{-1}\left[\frac{3 n (t-t_{0})}{2 K_{\tau}}\right]\Bigg\},
\end{aligned}
\end{equation}
\begin{equation}\label{eq:c_BI_2_1}
\begin{aligned}
    c(t) =& c_{1} \left[\frac{4K_{\tau}^2}{9n^{2}} -(t-t_{0})^2\right]^{\frac{1}{3 n}}\\
    &\exp \Bigg\{-m_{3}\tanh ^{-1}\left[\frac{3 n (t-t_{0})}{2K_{\tau}}\right]\Bigg\},
\end{aligned}
\end{equation}
where
\begin{gather}
    m_{1} = \frac{2 (k_{ab}+k_{ac})}{3 n \Omega},\\
    m_{2} = \frac{2 (k_{ac}-2 k_{ab})}{3 n \Omega},\\
    m_{3} = \frac{2 (k_{ab}-2 k_{ac})}{3 n \Omega}.
\end{gather}
In Fig. \ref{fig:BI_example2_1} we show an example of the evolution of scale factors: at the beginning, the Universe is spatially one dimensional (and therefore singular) and becomes spatially one dimensional again after a process of expansion and contraction, as described by the behavior of $\tau$.
Hence, the relative differences between the scale factors are greatest at the beginning and the end of the cosmic history, with a time interval in which the values of the scale factors are close to each other. 

\begin{figure}[ht]
    \begin{subfigure}[ht]{\linewidth}
        \includegraphics[width=8.6cm, height=4.5cm]{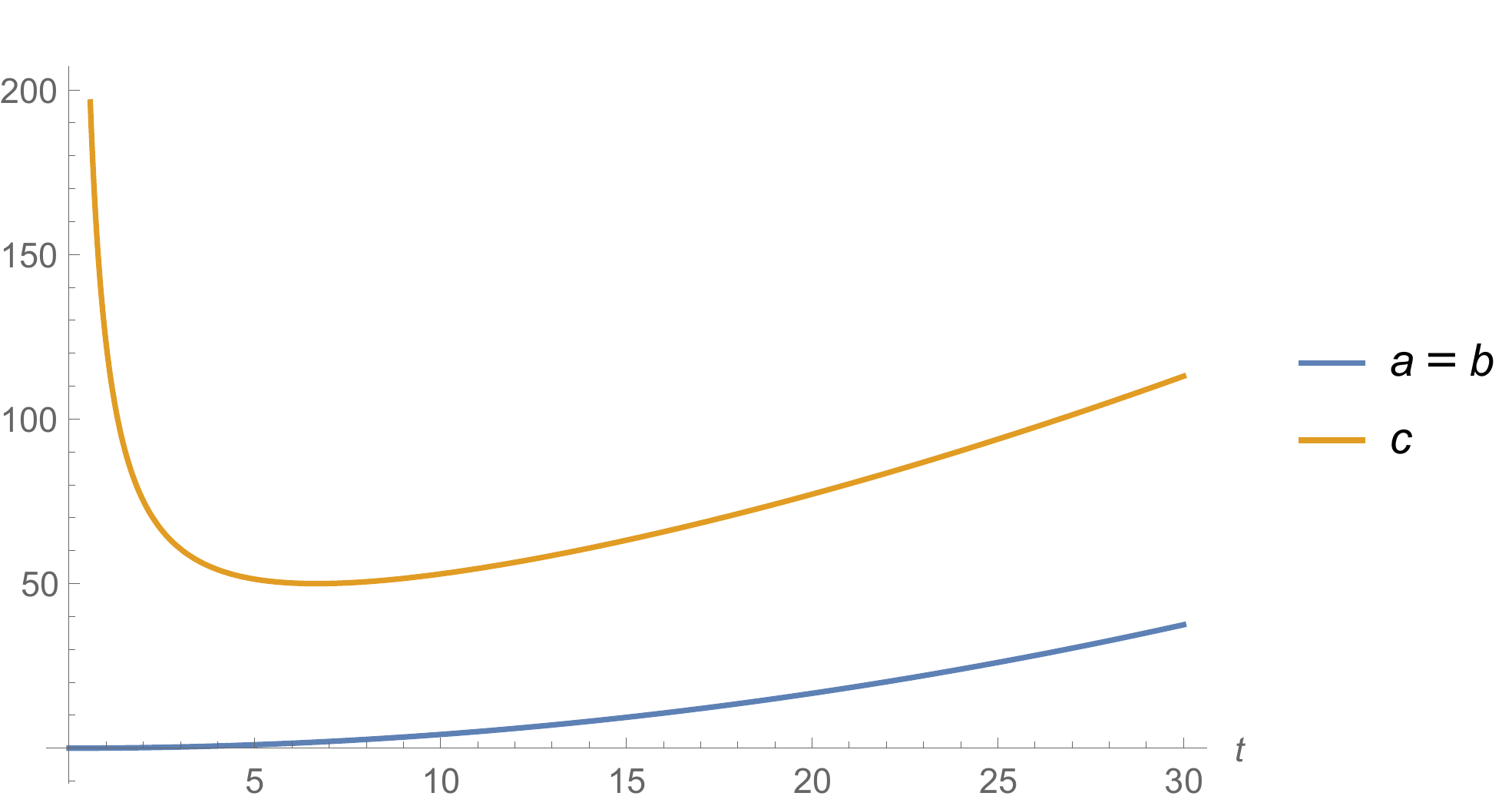}
        \caption{Scale factors}
        \label{fig:BI_example2_1__b_a}
    \end{subfigure}
    \begin{subfigure}[ht]{\linewidth}
        \includegraphics[width=8.6cm, height=4.5cm]{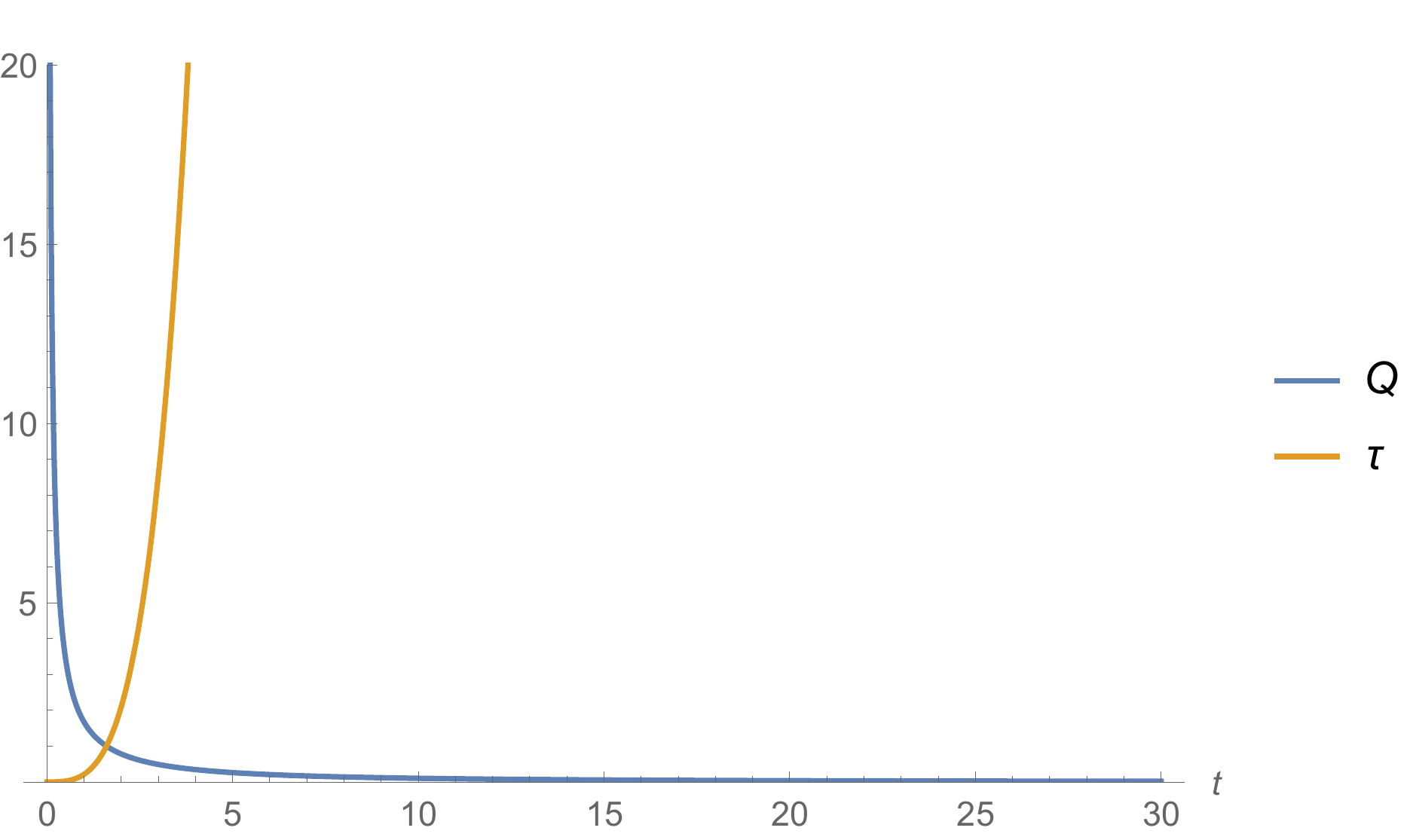} 
        \caption{Non-metricity scalar and $\tau$.}
        \label{fig:BI_example2_1_b_b}
    \end{subfigure}
    \caption{Evolution of \eqref{eq:tau_BI_2_1_b}-\eqref{eq:c_BI_2_1_b} with  values $n=\frac{1}{3}$, $K_{\tau}=\frac{10}{3}$, $\mathcal{Q}_{0}=1$, $a_{1}=c_{1}=\frac{1}{24}$, $m_{1}=1$, $m_{2}=2$, and $t_{0}=-\frac{20}{3}$.}
    \label{fig:BI_example2_1_2}
\end{figure}

On the other hand, if 
\begin{equation}
    \tau =  \left(\frac{\mathcal{Q}_{0}}{\mathcal{Q}}\right)^{\frac{1}{n}},
\end{equation}
with $n$ still an odd integer or a rational number with an odd denominator, the solution of Eq. \eqref{eq:exampleBI_2_2} is
\begin{equation}\label{eq:tau_BI_2_1_b}
    \tau(t) = a_{1}b_{1}c_{1} \left[(t-t_{0})^2 - \frac{4K_{\tau}^2}{9n^{2}}\right]^{\frac{1}{n}},
\end{equation}
Given Eq. \eqref{eq:tau_BI_2_1_b}, and choosing two scale factors proportional to each other, e.g. $a=b$, we have
\begin{equation}\label{eq:a_BI_2_1_b}
\begin{aligned}
    a(t) =  a_{1} \left[(t-t_{0})^2 - \frac{4K_{\tau}^2}{9n^{2}}\right]^{\frac{1}{3 n}}\left[ \frac{3n\left( t - t_{0} \right)-2 K_{\tau}}{3n\left( t - t_{0} \right) + 2 K_{\tau}} \right]^{m_{1}},
\end{aligned}
\end{equation}
and
\begin{equation}\label{eq:c_BI_2_1_b}
\begin{aligned}
   c(t) = c_{1} \left[(t-t_{0})^2 - \frac{4K_{\tau}^2}{9n^{2}}\right]^{\frac{1}{3 n}} \left[\frac{3n\left( t - t_{0} \right)-2 K_{\tau}}{3n\left( t - t_{0} \right) + 2 K_{\tau}} \right]^{-m_{2}},
\end{aligned}
\end{equation}
where
\begin{gather}
    m_{1} = \frac{1}{3 n},\\
    m_{2} = \frac{2}{3 n}.
\end{gather}
A representation of the cosmological evolution is given in Fig. \ref{fig:BI_example2_1_2}. After an initial singular phase, in which the Universe is spatially one dimensional, the scale factors grow showing a similar behavior for large values of $t$. This trend can be immediately verified by taking the limit for $t\rightarrow\infty$ of Eq. \eqref{eq:a_BI_2_1_b} and of Eq. \eqref{eq:c_BI_2_1_b}. 
 
\subsubsection{Case: \texorpdfstring{$n=2\bar{w}$}{}}\label{subsec:BI_Example_2_2}
\begin{figure}[ht]
    \begin{subfigure}[ht]{\linewidth}
        \includegraphics[width=8.6cm, height=4.5cm]{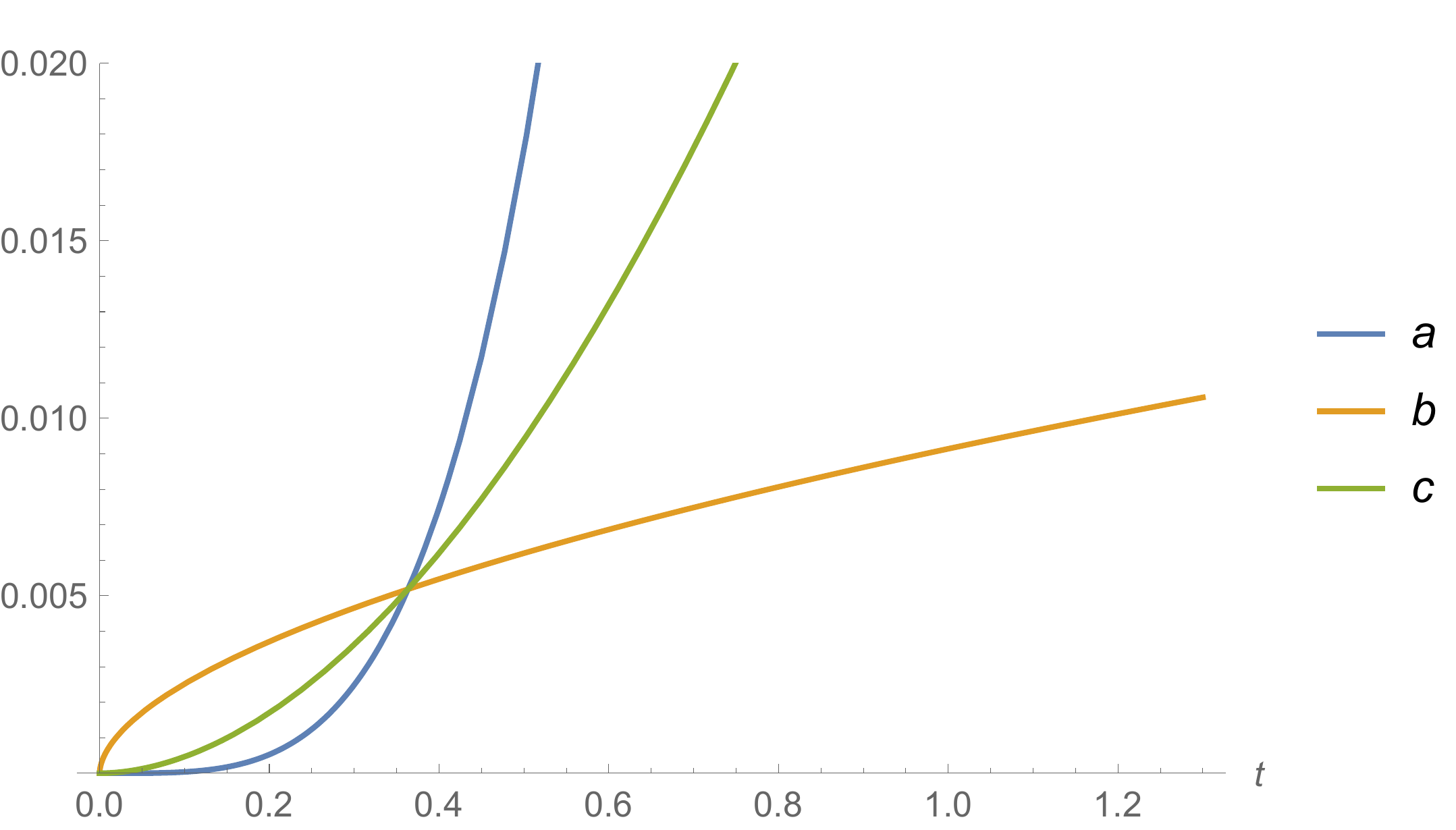} 
        \caption{Scale factors}
        \label{fig:BI_example2_2_a}
    \end{subfigure}
    \begin{subfigure}[ht]{\linewidth}
        \includegraphics[width=8.6cm, height=4.5cm]{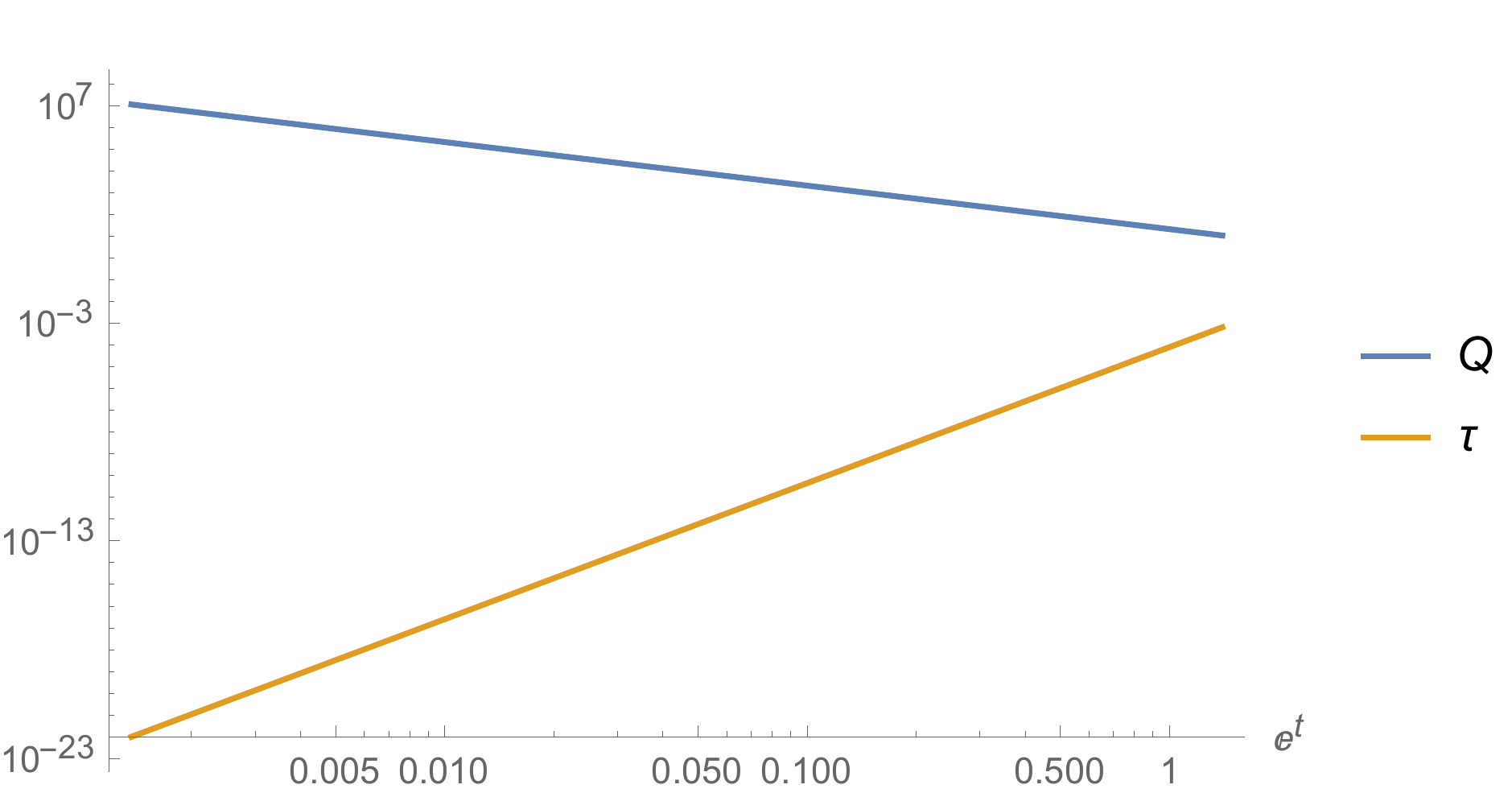}
        \caption{Double logarithm plot of non-metricity scalar and $\tau$.}
        \label{fig:BI_example2_2_b}
    \end{subfigure}
    \caption{Evolution of \eqref{eq:tau_BI_2_2}-\eqref{eq:c_BI_2_2} with  values $w=0.16$, $K_{\tau}=1.25$, $\mathcal{Q}_{0}=1$, $a_{1}=0.25$, $b_{1}=9.14 \cdot 10^{-3}$, $c_{1}=3.42 \cdot 10^{-2}$, $m_{1}=1.74$, $m_{2}=-1.52$, $m_{3}=-0.22$, and $t_{0}=0$.}
    \label{fig:BI_example2_2}
\end{figure}

We now set $n=2\bar{w}$, $\bar{w} \neq 0$, and 
\begin{equation}
    \tau = \left(\frac{\mathcal{Q}_{0}}{\mathcal{Q}}\right)^{\frac{1}{2\bar{w}}}.
\end{equation}
In this case, the solution of Eq. \eqref{eq:exampleBI_2_2} is
\begin{equation}\label{eq:tau_BI_2_2}
    \tau(t) = a_{1}b_{1}c_{1}\left(t + t_{0}\right)^{\frac{1}{\bar{w}}}.
\end{equation}  
The scale factors are
\begin{equation}\label{eq:a_BI_2_2}
a(t) = a_{1}\left(t + t_{0}\right)^{\frac{1}{3 \bar{w}}+m_{1}},
\end{equation}
\begin{equation}\label{eq:b_BI_2_2}
    b(t) = b_{1} \left(t + t_{0}\right)^{\frac{1}{3 \bar{w}}+m_{2}},
\end{equation}
\begin{equation}\label{eq:c_BI_2_2}
    c(t) = c_{1} \left(t + t_{0}\right)^{\frac{1}{3 \bar{w}}+m_{3}},
\end{equation}
where  
\begin{gather}
    m_{1} = \frac{K_{\tau}\sqrt{\mathcal{Q}_{0}} (k_{ab}+k_{ac})}{3 \bar{w} \Omega \sqrt{\mathcal{Q}_{0}K_{\tau}^2+6}},\\
    m_{2} = \frac{K_{\tau}\sqrt{\mathcal{Q}_{0}} (k_{ac}-2 k_{ab})}{3 \bar{w} \Omega \sqrt{\mathcal{Q}_{0}K_{\tau}^2+6}},\\
    m_{3} = \frac{K_{\tau}\sqrt{\mathcal{Q}_{0}} (k_{ab}-2 k_{ac})}{3 \bar{w} \Omega \sqrt{\mathcal{Q}_{0}K_{\tau}^2+6}}.
\end{gather}    
Note that the sum of the exponents is equal to $1/\bar{w}$ since
\begin{equation}
    m_{1} + m_{2} + m_{3} = 0.
\end{equation}
With $t_{0}=0$, we recover the results of Sec. \ref{subsec:BI_Example_1}, where the scale factors are
\begin{equation}
    a(t) = a_{1} t^{\frac{1}{3\bar{w}}+m_{1}}, \quad 
    b(t) = b_{1} t^{\frac{1}{3\bar{w}}+m_{2}}, \quad
    c(t) = c_{1} t^{\frac{1}{3\bar{w}}+m_{3}}.
\end{equation}
We give a representation of the evolution of the system in Fig. \ref{fig:BI_example2_2}. The scale factors are always growing, but their growth rate is different. We notice a particular instant of time where the scale factors coincide.

\subsubsection{Case: \texorpdfstring{$n=\bar{w}$}{}}\label{subsec:BI_Example_2_3}
\begin{figure}[ht]
    \centering
    \begin{subfigure}[ht]{\linewidth}
        \centering
        \includegraphics[width=8.6cm, height=4.5cm]{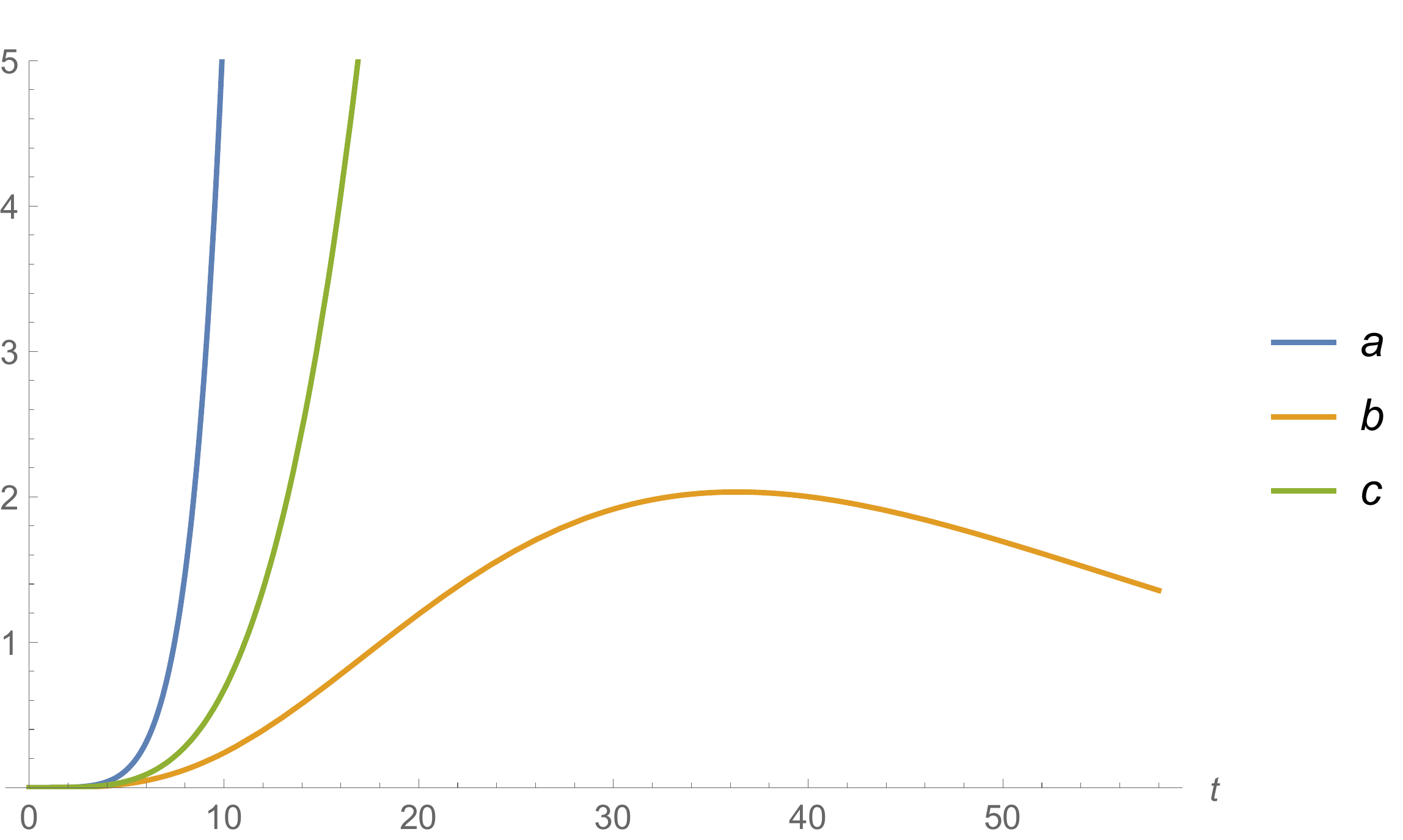} 
        \caption{Scale factors}
        \label{fig:BI_example2_3_a}
    \end{subfigure}
    \begin{subfigure}[ht]{\linewidth}
        \centering
        \includegraphics[width=8.6cm, height=4.5cm]{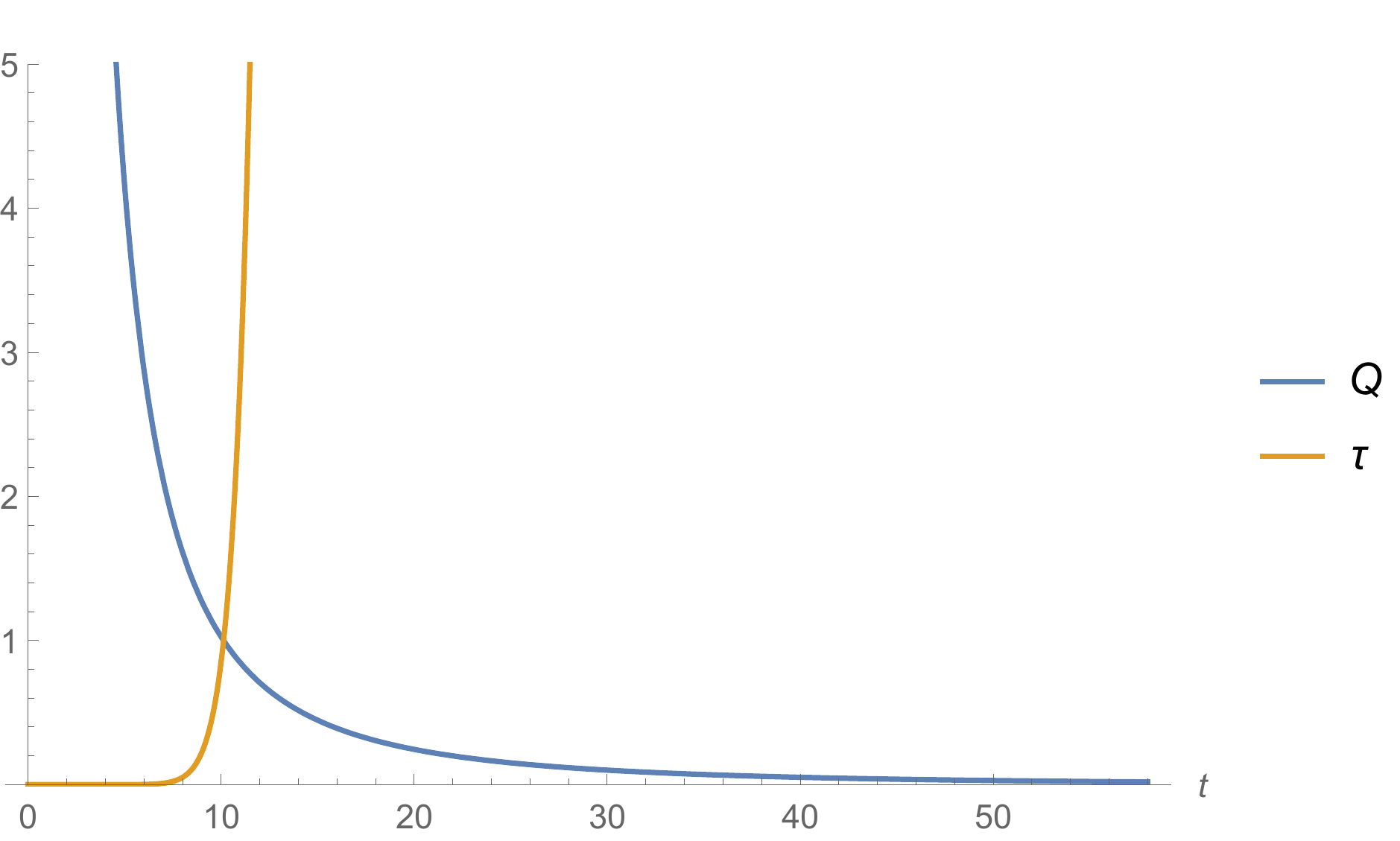}
        \caption{Non-metricity scalar and $\tau$.}
        \label{fig:BI_example2_3_b}
    \end{subfigure}
    \caption{Evolution of \eqref{eq:tau_BI_2_3}-\eqref{eq:c_BI_2_3} with  values $\bar{w}=0.16$, $K_{\tau}=0.55$, $\mathcal{Q}_{0}=1$, $a_{1}=b_{1}=c_{1}=515$, $m_{1}=0.17$, $m_{2}=-0.14$, $m_{3}=3.45 \cdot 10^{-2}$, and $t_{0}=0$.}
    \label{fig:BI_example2_3}
\end{figure}

If $n=\bar{w}$, $\bar{w} \neq 0$, and
\begin{equation}
    \tau = \left(\frac{\mathcal{Q}_{0}}{\mathcal{Q}}\right)^{\frac{1}{\bar{w}}},
\end{equation}
then the solution of Eq. \eqref{eq:exampleBI_2_2} is
\begin{equation}\label{eq:tau_BI_2_3}
    \tau(t) = a_{1}b_{1}c_{1} \bigg\{\sinh \left[\frac{1}{4} K_{\tau}\mathcal{Q}_{0} \bar{w} (t+t_{0})\right]\bigg\}^{\frac{2}{\bar{w}}}.
\end{equation}
The scale factors are
\begin{equation}\label{eq:a_BI_2_3}
\begin{aligned}
    a(t) = a_{1} \bigg\{\sinh \left[\frac{1}{4} K_{\tau}\mathcal{Q}_{0} \bar{w} (t+t_{0})\right]\bigg\}^{\frac{2}{3 \bar{w}}} e^{m_{1}t},
\end{aligned}
\end{equation}
\begin{equation}\label{eq:b_BI_2_3}
\begin{aligned}
    b(t) = b_{1} \bigg\{\sinh \left[\frac{1}{4} K_{\tau} \mathcal{Q}_{0} \bar{w} (t+t_{0})\right]\bigg\}^{\frac{2}{3 \bar{w}}} e^{m_{2}t},
\end{aligned}
\end{equation}
\begin{equation}\label{eq:c_BI_2_3}
    c(t) = c_{1} \bigg\{\sinh\left[\frac{1}{4} K_{\tau} \mathcal{Q}_{0} \bar{w} (t+t_{0})\right]\bigg\}^{\frac{2}{3 \bar{w}}} e^{m_{3} t},
\end{equation}
where
\begin{gather}
    m_{1} = \frac{K_{\tau} \mathcal{Q}_{0} (k_{ab}+k_{ac})}{6 \Omega},\\
    m_{2} = \frac{K_{\tau} \mathcal{Q}_{0} (k_{ac}-2 k_{ab})}{6 \Omega},\\
    m_{3} = \frac{K_{\tau} \mathcal{Q}_{0} (k_{ab}-2 k_{ac})}{6 \Omega}.
\end{gather}
Let us consider the example shown in Fig. \ref{fig:BI_example2_3} where two scale factors grow up to infinity while the third one,  which in the figure is represented by the scale factor $b$, once it has reached a maximum, tends to zero. Therefore, during its evolution the Universe undergoes a first phase where the scale factors are all growing and a second one where the spacetime tends to become spatially two dimensional and therefore singular.

\section{Reconstruction method: FLRW}\label{ch:recontruction_FRW}
In this section, we will consider spatially flat FLRW cosmologies. As it will be evident, the higher symmetry of the FLRW spacetime, compared to BI, will make it easier to apply the reconstruction method. 

\subsection{Example 1: reconstruction from a time dependent scale factor}
\label{subsec:FLRW_Example_1}
To start with, we take two scale factors $a=a(t)$ into account: a power law and an exponential function of time.
\subsubsection{Scale factor as a power law}\label{subsec:FLRW_Example_1_1}
Setting
\begin{equation}\label{atn}
    a(t)=a_{0}t^{n},
\end{equation}
with $a_{0}$ a dimensional constant, the non-metricity scalar assumes the form
\begin{equation}\label{V_Q1}
    \mathcal{Q}=6H^{2}=6n^{2}t^{-2}.
\end{equation}
Inverting relation \eqref{V_Q1}, we may express the scale factor and density $\rho$ as function of $\mathcal{Q}$:
\begin{equation}
    a(\mathcal{Q}) = a_{0} \left( \frac{\alpha}{\mathcal{Q}} \right)^{\frac{n}{2}} 
\end{equation}
and
\begin{equation}\label{V_rho1}
    \rho = \rho_{0} a^{-3(1+\bar{w})} = \rho_{0}a_{0}^{-3(1+\bar{w})}\left(
    \frac{\mathcal{Q}}{\alpha} \right)^{\frac{3}{2}n(1+\bar{w})},
\end{equation}
with $\alpha=6n^{2}$. Replacing Eq. \eqref{V_rho1} in Eq. \eqref{eq:cosmology_FLRW_1} and solving the resulting differential equation, we obtain the function
\begin{equation}\label{V_f1}
    f(\mathcal{Q}) = f_{0} \sqrt{\mathcal{Q}} + f_{1}\mathcal{Q}^{\frac{3}{2}(1+\bar{w})n}
\end{equation}
with 
\begin{equation}
f_{1} = \frac{2 \rho_{0}}{3n(1+\bar{w})-1}a_{0}^{-3(1+\bar{w})}\alpha^{-\frac{3}{2}(1+\bar{w})n}.
\end{equation}
The function \eqref{V_f1} is similar to the ones that have been mostly used in literature so far \cite{BeltranJimenez:2017tkd, Jim_nez_2020}\footnote{Notice that if we were to start with $n=\frac{2}{3(1+\bar{w})}$, the solution that encompasses all the classical Friedmannian cosmological solutions, then the \eqref{V_f1} would give $f(\mathcal{Q})\propto\mathcal{Q}$. This result implies that $f(\mathcal{Q})$ gravity can have, at most, one cosmological solution in common with GR.}.

Now we will show that, as anticipated in Sec. \ref{subsec:BI_Example_1}, the solution Eq. \eqref{V_f1} is valid even if we choose to resolve the cosmological equations in the case of fluids with $w \neq\bar{w}$. For example, let us consider $\bar{w}=0$, then
\begin{equation}\label{test_f(Q)}
    f(\mathcal{Q}) = f_{0}\sqrt{\mathcal{Q}} + \frac{2 \rho_{0}}{3n-1}a_{0}^{-3} \left(\frac{\mathcal{Q}}{\alpha}\right)^{\frac{3}{2}n},
\end{equation}
and a fluid with $w=\frac{1}{3}$, i.e. radiation, for which 
\begin{equation}\label{test_rho}
    \rho_{1} = \rho_{1,0} a^{-4}.
\end{equation}
Substituting Eqs. \eqref{test_f(Q)} and \eqref{test_rho} into Eq. \eqref{eq:cosmology_FLRW_1}, we obtain the following expression for $a(t)$:
\begin{equation}\label{test_scale_factor}
    a(t) = \left(\frac{2 \sqrt[3]{\alpha }}{3}\right)^{\frac{9 n}{8}} \sqrt[4]{\frac{a_{0}^3 \rho_{1,0}}{\rho_{0}}} \left(\frac{t+t_{0}}{n}\right)^{\frac{3 n}{4}},
\end{equation}
which, for the same value of $n$, is clearly different from \eqref{atn}.

\subsubsection{Scale factor as an exponential function}\label{subsec:FLRW_Example_1_2}
\begin{figure}[ht]
    \centering
    \begin{subfigure}[ht]{\linewidth}
        \centering
        \includegraphics[width=8.6cm, height=4.5cm]{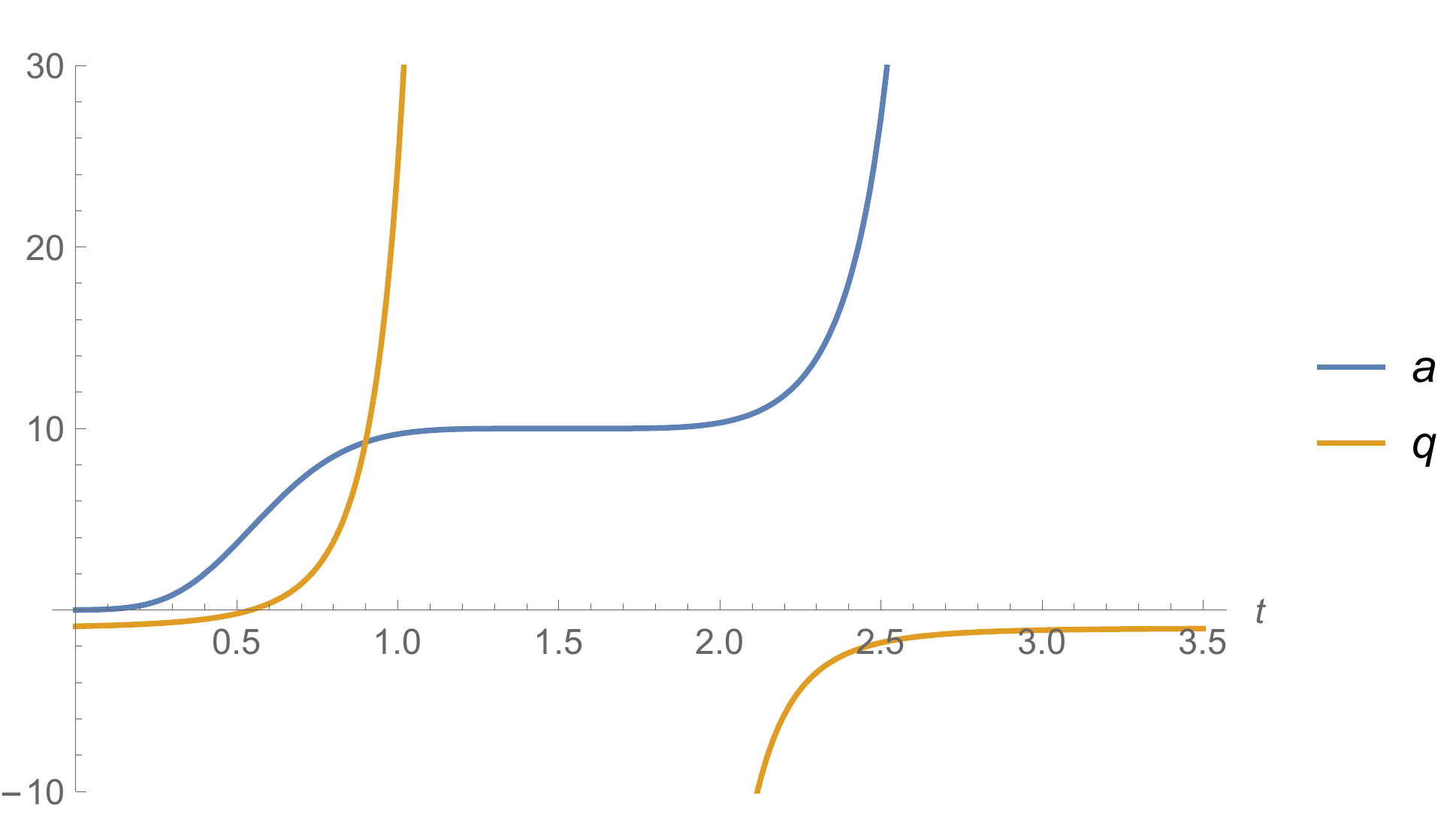} 
        \caption{Scale factor and deceleration parameter}
        \label{fig:FLRW_Example1_1}
    \end{subfigure}
    \begin{subfigure}[ht]{\linewidth}
        \centering
        \includegraphics[width=8.6cm, height=4.5cm]{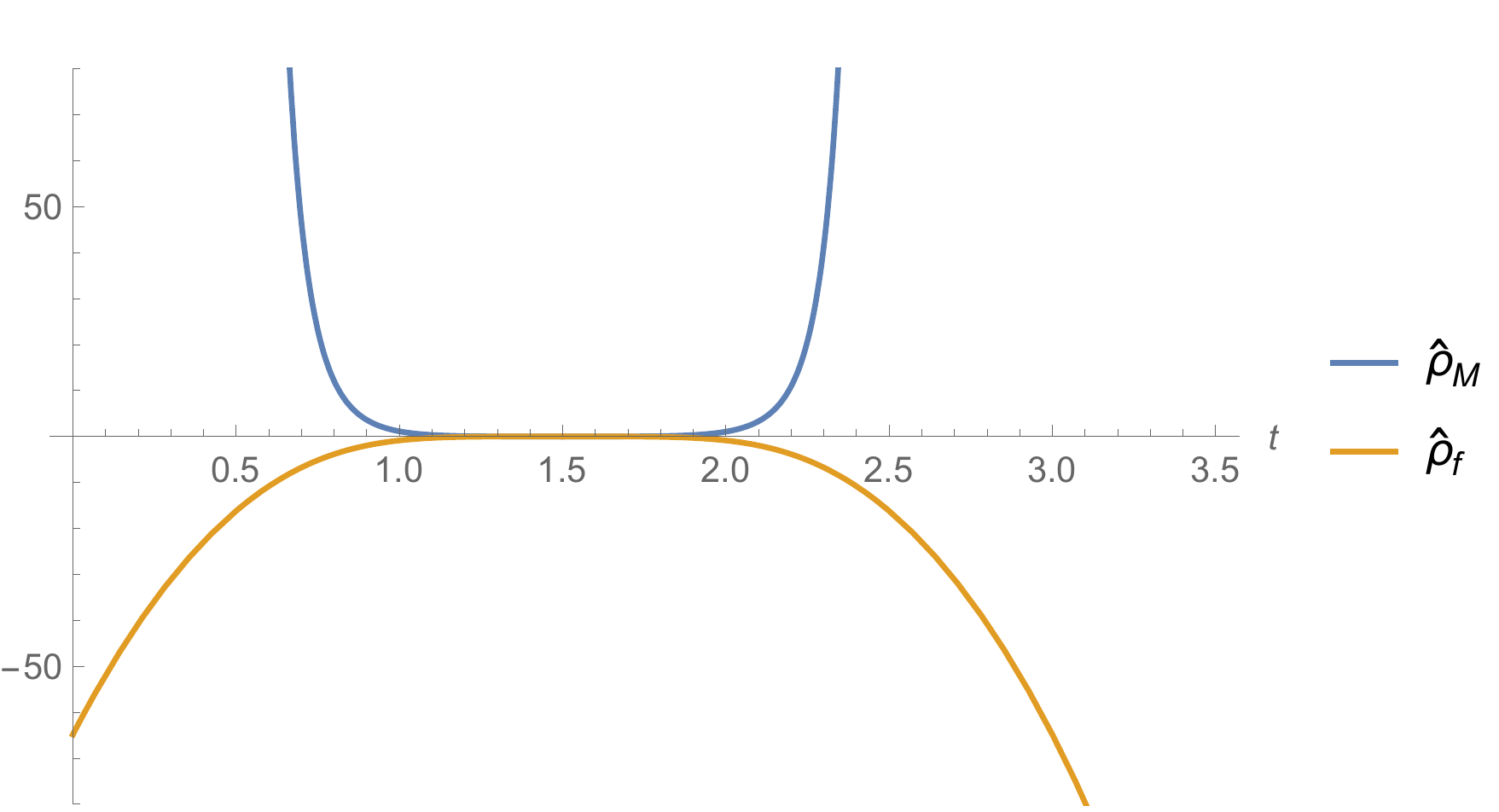} 
        \caption{$\hat{\rho}_{M}$ and $\hat{\rho}_{f}$}
        \label{fig:FLRW_Example1_2}
    \end{subfigure}
    \caption{Evolution of \eqref{scFRLW2} with  values $n=2$, $m=1$, $\rho_{0}=1$, $a_{0}=10$, $f_{0}=0$, $t_{0}=\frac{3}{2}$, and $\bar{w}=0$.}
    \label{fig:FLRW_Example1}
\end{figure}

We set now
\begin{equation}\label{scFRLW2}
    a(t)=a_{0}e^{m(t-t_{0})^{2n+1}},
\end{equation}
where $m>0$, $a_{0}$ and $t_{0}$ are generic constants, and $n$ is a natural number.
The scale factor \eqref{scFRLW2} describes a cosmic scenario in which all of the three main phases of the cosmological evolution (inflation, Friedmann phase, and dark phase) are represented (see Fig. \ref{fig:FLRW_Example1_1}).  The duration of the Friedmann phase is related to the value of the odd exponent $2n+1$, which therefore plays a crucial role. 
The non-metricity scalar associated with the scale factor \eqref{scFRLW2} has the form
\begin{equation}\label{V_Q2}
    \mathcal{Q} = 6(2n+1)^{2}m^{2}(t-t_{0})^{4n}.
\end{equation}
Again, inverting Eq. \eqref{V_Q2}, we derive the expressions of the scale factor $a$ and the density $\rho$ as function of $\mathcal{Q}$:
\begin{equation}
    a(\mathcal{Q}) = a_{0} \exp\left[ m\left(\frac{\mathcal{Q}}{\alpha}\right)^{\frac{2n+1}{4n}}\right]
\end{equation}
and
\begin{equation}\label{V_rho2}
    \rho = \rho_{0} a_{0}^{-3(1+\bar{w})}\exp\left[-3m(1+\bar{w}) \left(\frac{\mathcal{Q}}{\alpha}\right)^{\frac{2n+1}{4n}}\right],
\end{equation}
with $\alpha=6(2n+1)^{2}m^{2}$. 

Inserting Eq. \eqref{V_rho2} into Eq. \eqref{eq:cosmology_FLRW_1} and solving, we get the function
\begin{equation}
    f(\mathcal{Q}) = f_{0} \sqrt{\mathcal{Q}} - f_{1} \mathcal{Q}^{\frac{1}{2}} \Gamma \left[-\frac{2 n}{2 n+1},3m(1+\bar{w})\left(\frac{\mathcal{Q}}{\alpha}\right)^{\frac{2n+1}{4n}}\right]
\end{equation}
where $\Gamma$ is the incomplete gamma function, and
\begin{equation}
    f_{1}=\frac{4n }{2n+1} \frac{\rho_{0}}{\alpha^{2}} a_{0}^{-3(1+\bar{w})} \left[ 3m(1+\bar{w})  \right]^{\frac{2 n}{2 n+1}}.
\end{equation}
As it can be seen in Fig. \ref{fig:FLRW_Example1_2}, here the effective gravitational constant $1/f'(\mathcal{Q})$ is positive, whereas the non-metricity correction $\hat{\rho}_f$ is always negative. However, $\hat{\rho}_f$ grows slower than the matter term $\hat{\rho}_M$, so the scale factor will always tend to increase, and the cosmology expands.

\subsection{Example 2: reconstruction from the time derivative of the scale factor}\label{subsec:FLRW_Example_2}
The previous examples relied explicitly on the inversion of the expression of the non--metricity scalar, i.e. we always needed to obtain $t=t(\mathcal{Q})$. It is clear that such inversion is not always possible analytically.  Another option is to give an implicit expression for the scale factor. In particular, we can consider the scale factor $a(t)$ as defined by a suitable differential equation,
\begin{equation}
    \dot{a}=h(a),
\end{equation}
with $h(a)$ a generic function of the scale factor. Then, we can express the non-metricity scalar in the form
\begin{equation}\label{eq:exampleFLRW_2_1}
    \mathcal{Q} = 6 \left[\frac{h(a)}{a}\right]^{2}.
\end{equation}
As an example, let us consider the relation
\begin{equation}
    \dot{a}=\frac{2\Omega}{\sqrt{\Lambda}}\sqrt{a-\Lambda a^{2}},
\end{equation}
where $\Omega$ and $\Lambda$ are generic constants,
from which we derive the evolution of $a(t)$.
\begin{equation}\label{scFRLW_2}
    a(t)=\frac{1}{\Lambda}\sin^{2}(\Omega t).
\end{equation}
Equation \eqref{eq:exampleFLRW_2_1} allows us to get the scale factor and density $\rho$ as a function of $\mathcal{Q}$,
\begin{equation}
    \frac{1}{a} = \Lambda + \frac{\Lambda \mathcal{Q}}{24\Omega^{2}}
\end{equation}
and
\begin{equation}\label{V_rhoB}
    \rho = \rho_{0} \left(\Lambda +\frac{\Lambda  \mathcal{Q}}{24 \Omega ^2} \right)^{3(1+\bar{w})}.
\end{equation}
Replacing Eq. \eqref{V_rhoB} in Eq. \eqref{eq:cosmology_FLRW_1} and solving, we find the solution,
\begin{equation}
\begin{aligned}
    f(\mathcal{Q}) =& f_{0} \sqrt{\mathcal{Q}} +  \, _2F_1\left[\frac{5}{2},-3 \bar{w} ;\frac{7}{2};-\frac{\mathcal{Q}}{24 \Omega ^2}\right] f_{1} \mathcal{Q}^3 +\\
    &+ \, _2F_1\left[\frac{3}{2},-3 \bar{w} ;\frac{5}{2};-\frac{\mathcal{Q}}{24 \Omega ^2}\right] f_{2} \mathcal{Q}^2 +\\
    &+ \, _2F_1\left[\frac{1}{2},-3 \bar{w} ;\frac{3}{2};-\frac{\mathcal{Q}}{24 \Omega ^2}\right]  f_{3} \mathcal{Q} +\\
    &- \, _2F_1\left[-\frac{1}{2},-3 \bar{w} ;\frac{1}{2};-\frac{\mathcal{Q}}{24 \Omega ^2}\right] f_{4},
\end{aligned}    
\end{equation}
where $ _{2}F_{1}$ denotes the hypergeometric function, and $f_{i}$ ($i=1,...,4$) are constants depending on $\Lambda$, $\Omega$, $\rho_{0}$, and $\bar{w}$.

\begin{figure}[ht]
    \centering
    \begin{subfigure}[ht]{\linewidth}
        \centering
        \includegraphics[width=8.6cm, height=4.5cm]{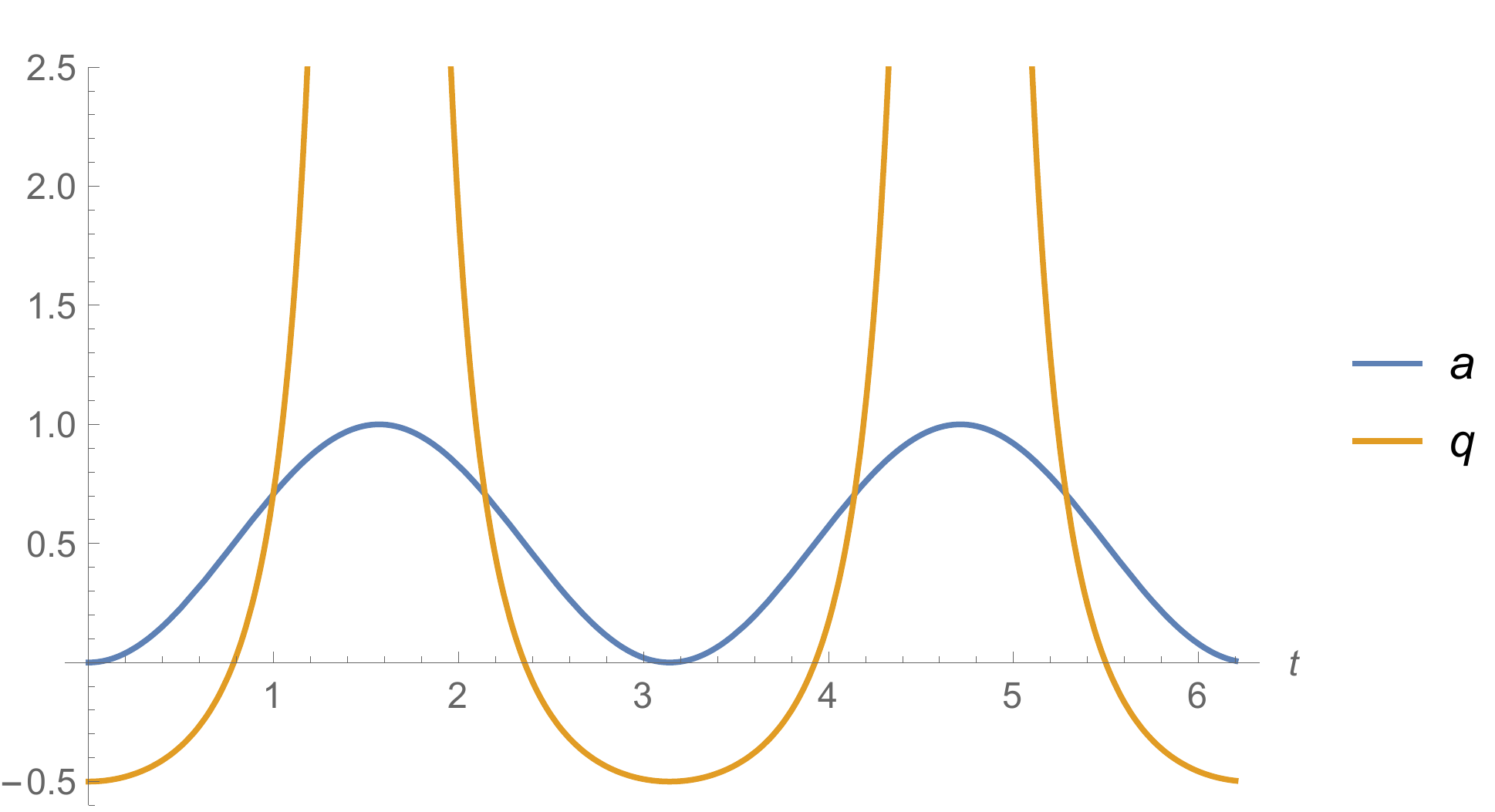}
        \caption{Scale factor and deceleration parameter}
        \label{fig:FLRW_Example2_1}
    \end{subfigure}
    \begin{subfigure}[ht]{\linewidth}
        \centering
        \includegraphics[width=8.6cm, height=4.5cm]{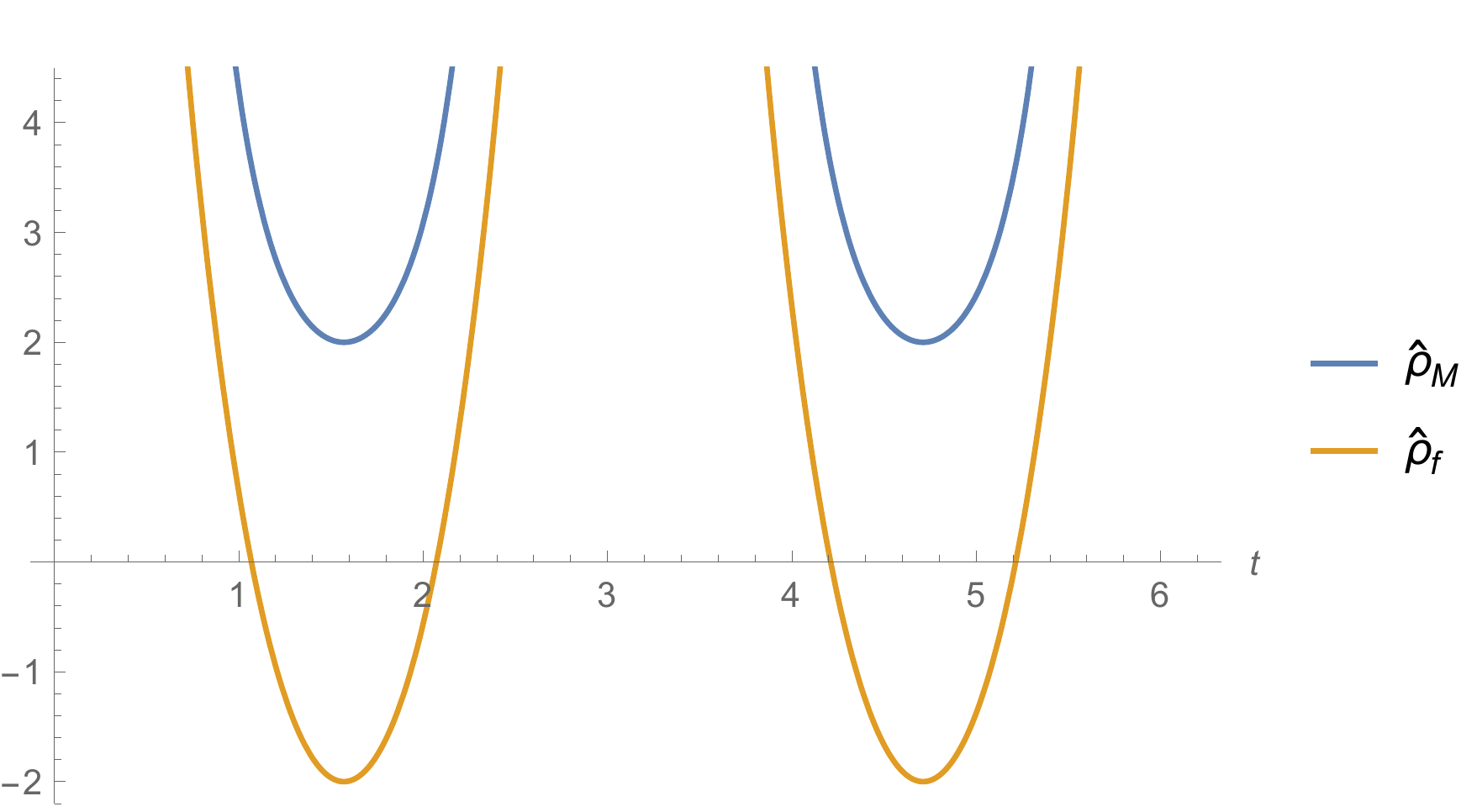}
        \caption{$\hat{\rho}_{M}$ and $\hat{\rho}_{f}$}
        \label{fig:FLRW_Example2_2}
    \end{subfigure}
    \caption{Evolution of \eqref{scFRLW_2} with  values $\Lambda=1$, $\Omega=1$, $\rho_{0}=1$, $f_{0}=0$, and $\bar{w}=0$.}
    \label{fig:FLRW_Example2}
\end{figure}

As we can see in Fig. \ref{fig:FLRW_Example2_1}, the scale factor \eqref{scFRLW_2} represents a cyclic universe in which every cycle is separated by a singularity. As in the previous example, the term $\hat{\rho}_M$ is always positive. However,  $\hat{\rho}_{f}$ changes sign. When  $\hat{\rho}_{f}<0$, the expansion slows up to the point in which the cosmology reaches an equilibrium and then starts contracting.  When $\hat{\rho}_{f}$  becomes positive, the contraction is slowed down up to the point in which the spacetime reaches the singularity with zero contraction rate but with positive acceleration. This fact suggests that the singularity might not be ``stable'' and, therefore, that one can use this solution for the analysis of pre-Big Bang scenarios in  $f(\mathcal{Q})$ gravity.

\subsection{Example 3: reconstruction from the deceleration parameter}\label{subsec:FLRW_Example_3}
Another way to avoid performing the inversion of the scale factor function is to express it indirectly in terms of a differential equation for the deceleration parameter. 
More specifically, we set 
\begin{equation}\label{eq:dec_par_FLRW_example3}
    \dot{q}=h(q),
\end{equation}
with $h(q)$ a generic function of the deceleration parameter. Equation \eqref{eq:dec_par_FLRW_example3} implies the following expression for the Hubble parameter and scale factor:
\begin{equation}
    \frac{1}{H(q)} = \int \frac{1+q}{h(q)}dq,
\end{equation}
\begin{equation}
    a(q)=\exp\left[\int\frac{H(q)}{h(q)}dq\right].
\end{equation}
As an example, let us consider the equation
\begin{equation}
    \dot{q}= q_{0}(1+q)\sqrt{q},
\end{equation}
where $q_{0}$ is a generic constant.
Using the above equations and remembering that $H^{2}=\mathcal{Q}/6$, we can write $q$ and $a$ as functions of $\mathcal{Q}$:
\begin{equation}
 q(\mathcal{Q})=\frac{3}{2}\frac{q_{0}^{2}}{\mathcal{Q}} ,
\end{equation}
\begin{equation}
\begin{split}
    a(\mathcal{Q}) = \sqrt{\frac{3q_{0}^{2}}{3q_{0}^{2} + 2\mathcal{Q}}}.
\end{split}
\end{equation}
Replacing in the Friedmann equation \eqref{eq:cosmology_FLRW_1},
\begin{equation}
\begin{aligned}
    f(\mathcal{Q}) =& f_{0} \sqrt{Q} + \frac{4}{3} \, _2F_1\left[\frac{1}{2},\frac{1}{2} (-3 \bar{w} -1);\frac{3}{2};-\frac{2 \mathcal{Q}}{3 q_{0}^2}\right] \frac{\rho_{0}}{q_{0}^{2}} \mathcal{Q} +\\
    &- 2 \,_2F_1\left[-\frac{1}{2},\frac{1}{2} (-3 \bar{w} -1);\frac{1}{2};-\frac{2 \mathcal{Q}}{3 q_{0}^2}\right] \rho_{0}.
\end{aligned}
\end{equation}
The definition of $\mathcal{Q}$, combined with the definition of $q$, gives a differential equation for $a(t)$ from which we derive
\begin{equation}\label{eq:a_FLRW_3}
    a(t)=a_{0} \sin \left[\frac{q_{0}}{2}\left(t-t_{0}\right)\right] + a_{1} \cos \left[\frac{q_{0}}{2}\left(t-t_{0}\right)\right],
\end{equation}
with the condition $ a_{0}^{2}+a_{1}^{2}=1$. 

\begin{figure}[ht]
    \centering
    \begin{subfigure}[ht]{\linewidth}
        \centering
        \includegraphics[width=8.6cm, height=4.5cm]{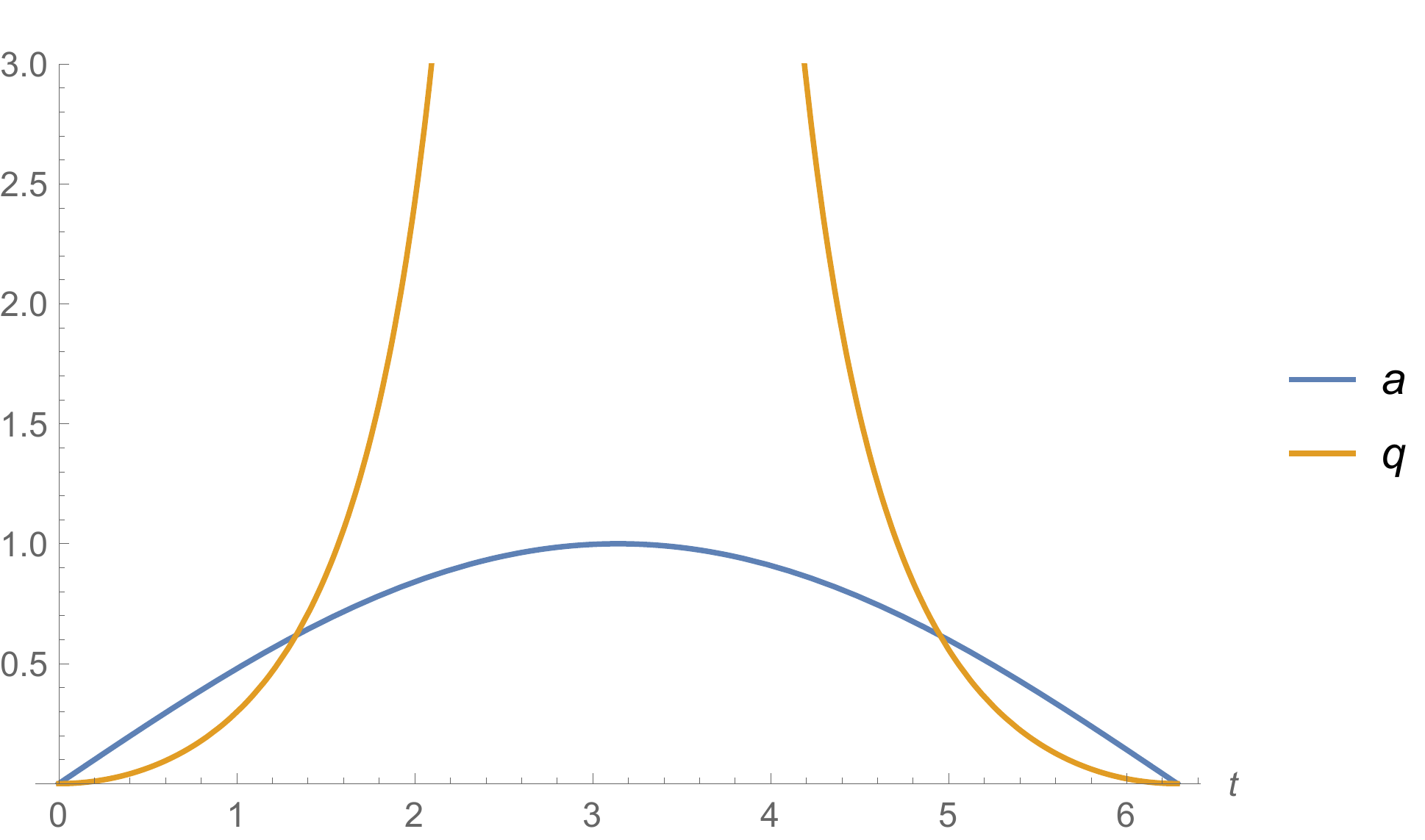} 
        \caption{Scale factor and deceleration parameter}
        \label{fig:FLRW_Example3_1}
    \end{subfigure}
    \begin{subfigure}[ht]{\linewidth}
        \centering
        \includegraphics[width=8.6cm, height=4.5cm]{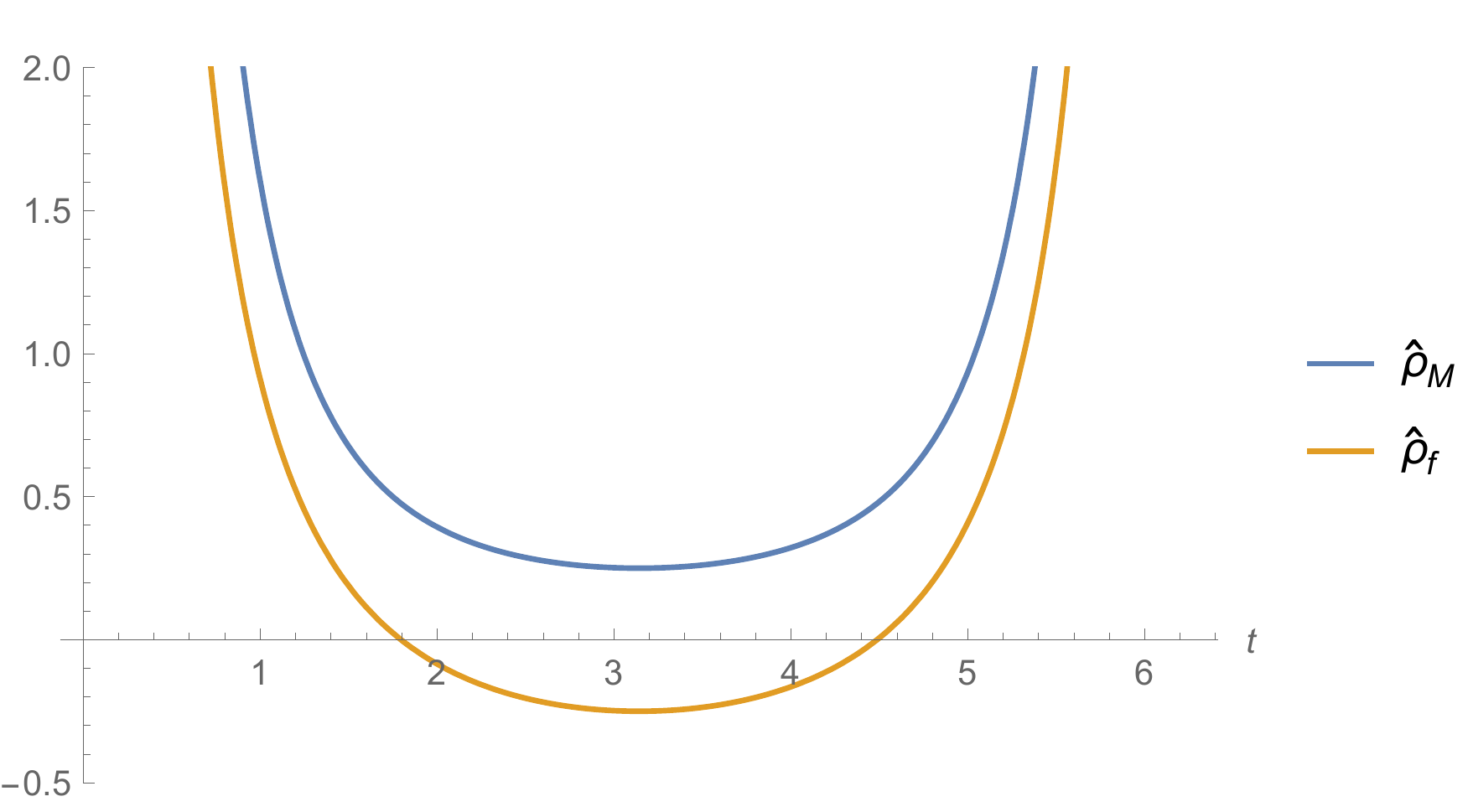} 
        \caption{$\hat{\rho}_{M}$ and $\hat{\rho}_{f}$}
        \label{fig:FLRW_Example3_2}
    \end{subfigure}
    \caption{Evolution of \eqref{eq:a_FLRW_3} with  values $a_{0}=a_{1}=\frac{1}{\sqrt{2}}$, $q_{0}=1$, $\rho_{0}=1$, $f_{0}=0$, $t_{0}=\frac{\pi}{2}$, and $\bar{w}=0$.}
    \label{fig:FLRW_Example3}
\end{figure}

As the above solution can be negative, we will limit ourselves to study only the first half-period (Fig. \ref{fig:FLRW_Example3}). It is clear that this solution represents again a universe enclosed between two singularities as it happens in Sec. \ref{subsec:FLRW_Example_2}. The difference is that departure and approach to the initial and final singularities happens with an expansion/contraction velocity different from zero. The behavior of non-metricity terms is similar to that in the previous subsection.

\subsection{Example 4: reconstruction from the time derivative of non-metricity scalar}\label{subsec:FLRW_Example_4}
\begin{figure}[ht]
    \centering
    \begin{subfigure}[ht]{\linewidth}
        \centering
        \includegraphics[width=8.6cm, height=4.5cm]{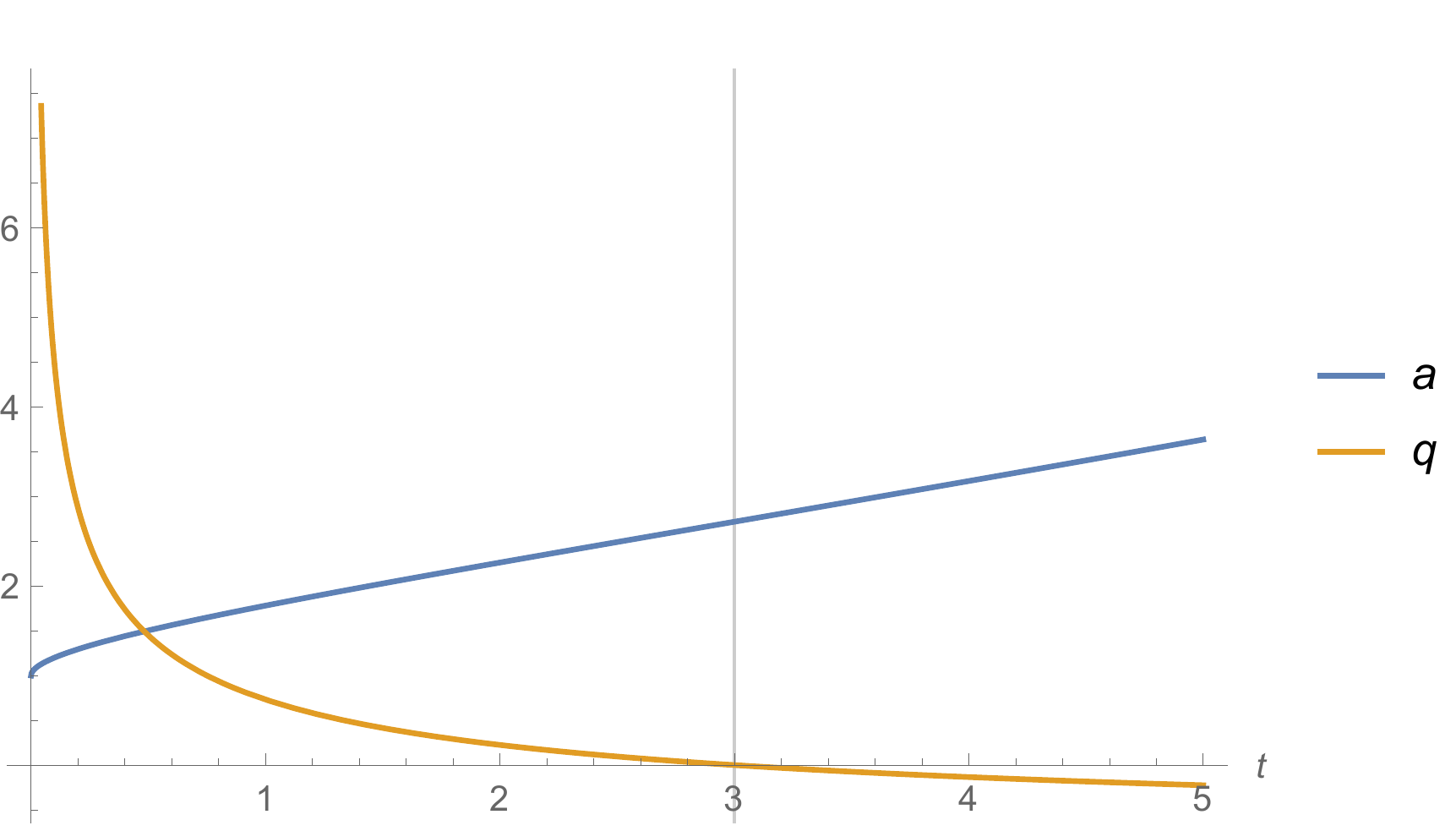} 
        \caption{Scale factor and deceleration parameter}
        \label{fig:FLRW_Example4_1}
    \end{subfigure}
    \begin{subfigure}[ht]{\linewidth}
        \centering
        \includegraphics[width=8.6cm, height=4.5cm]{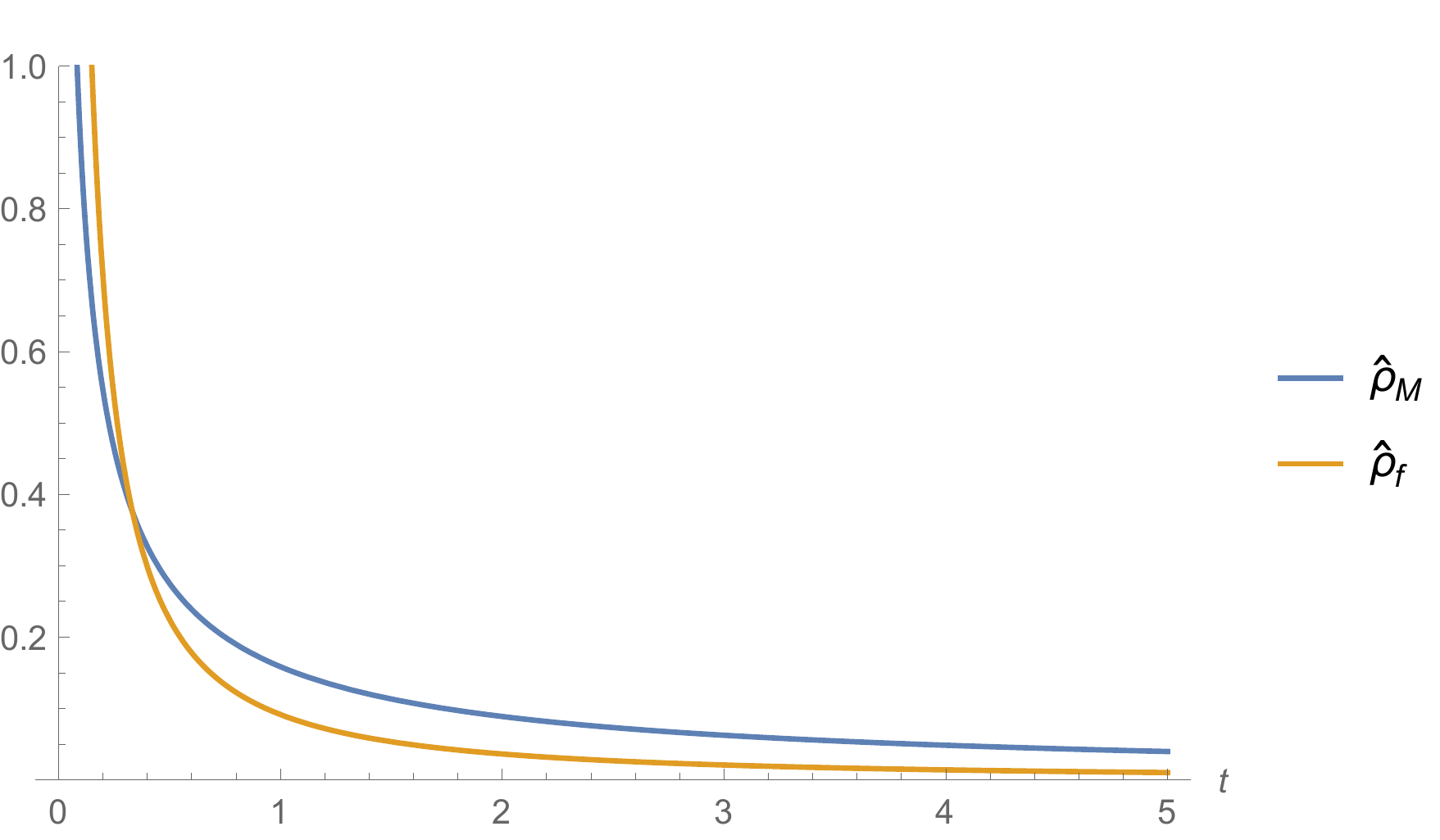} 
        \caption{$\hat{\rho}_{M}$ and $\hat{\rho}_{f}$}
        \label{fig:FLRW_Example4_2}
    \end{subfigure}
    \caption{Evolution of \eqref{V_Da} with values $n=2$, $\alpha=2$, $\rho_{0}=1$, $a_{0}=1$, $f_{0}=0$, $t_{0}=0$, and $\bar{w}=0$. The scale factor $a$ has an inflection point at $t=3$.}
    \label{fig:FLRW_Example4}
\end{figure}

In this last example, we reconstruct the scale factor and the function $f(\mathcal{Q})$ by imposing a differential constraint on the non-metricity scalar,
\begin{equation}\label{V_DQ}
    \dot{\mathcal{Q}}(t)=-\alpha \mathcal{Q}^{n}(t),
\end{equation}
with $\alpha$ an arbitrary constant.
From Eq. \eqref{V_DQ}, we get the solution
\begin{equation}\label{SolQEx4}
    \mathcal{Q}(t) = \left[\alpha(n-1) \left(t-t_{0}\right)\right]^{\frac{1}{1-n}}.
\end{equation}
By using the relation $\mathcal{Q}=6 H^{2}$, we first derive the scale factor as
\begin{equation}
    a( \mathcal{Q}) = a_{0} \exp\left[\sqrt{\frac{2}{3}}\frac{ \mathcal{Q}^{\frac{3}{2}-n}}{\alpha  (2 n-3)}\right]
\end{equation}
and then, using \eqref{SolQEx4},
\begin{equation}\label{V_Da}
    a(t) = a_{0}\exp \bigg\{ \sqrt{\frac{2}{3}}\frac{ \left[\alpha(n-1) (t-t_{0})\right]^{\frac{3-2n}{2-2 n}}}{\alpha  (2 n-3)}\bigg\}.
\end{equation}
The scale factor \eqref{V_Da} has an increasing trend for $n>\frac{3}{2}$ and $\alpha>0$ (see Fig. \ref{fig:FLRW_Example4_1}).
Equation \eqref{eq:cosmology_FLRW_1} gives, for $\bar{w}=0,1/3,1$
\begin{equation}
        f(\mathcal{Q}) = f_{0}\sqrt{\mathcal{Q}} + f_{1} \sqrt{\mathcal{Q}} \;\Gamma \left[\frac{1}{2 n-3},\frac{(1+\bar{w})\sqrt{6} \mathcal{Q}^{\frac{3}{2}-n}}{(2 n-3) \alpha }\right].
\end{equation}
where $f_{i}$ ($i=1,...,3$) are constants depending on $a_{0}$, $\alpha$, $\rho_{0}$, and $n$.

Looking at Fig. \ref{fig:FLRW_Example4_1}, it is evident that the scale factor changes concavity. Therefore, after a decelerated phase, the cosmology undergoes an accelerated expansion.  The term $\hat{\rho}_M$ is always positive; thus, the effective gravitational constant  $1/f'(\mathcal{Q})$ remains positive. The fact that the contribution of  $\hat{\rho}_f$ is also positive and that we are considering only standard fluids, suggests that non-metricity pressure $\hat{p}_f$  term in Eq. \eqref{eq:raychaudhuri_FLRW_effective} must be responsible for the accelerated expansion.

\section{Discussion and conclusions}\label{ch:conclusions}
In this paper, we derived and analyzed some exact cosmological solutions in the context of $f(\mathcal{Q})$ theory, with the aim to understand the role the non-metricity might play in the evolution of the Universe. 

The primary tools to perform this investigation were reconstruction techniques, in which the form of the scale factor(s) is assumed and the form of the function $f$ is recovered {\em a posteriori}. An essential step for developing the reconstruction algorithm is to reduce the cosmological equation to the simplest set of independent equations. In this respect, the relation between the different components of the Einstein equations is pivotal. We were able to show that, as it is well known in the case of FLRW metrics, also in the case of BI metrics one of the Einstein equations is dependent on the other when the conservation laws are taken into account. This feature has allowed us to reduce the number of equations to be solved in this case. In their original form, reconstruction techniques require some inversion of the scale factor or other related quantities. We have been able to go around this difficulty assigning a differential relation rather than an exact expression for the scale factors. This approach led to the derivation of several nontrivial solutions.

We started by studying the case of an anisotropic universe endowed with a BI metric. In this context, we found several solutions, such as universes where initial and final states are singular configurations with only one or two spatial dimensions (Secs. \ref{subsec:BI_Example_2_1} and \ref{subsec:BI_Example_2_3}), and more classical solutions where the scale factors are suitable power law functions (Secs. \ref{subsec:BI_Example_1} and \ref{subsec:BI_Example_2_2}). In Sec. \ref{subsec:BI_Example_2_1} we also obtained a universe that becomes more and more isotropic in the future.

In almost all of these solutions, we found that the non-metricity scalar presents some special features when the difference between the behavior of the scale factors (and hence anisotropy) has maximum or minimum. For example, in Fig. \ref{fig:BI_example2_1} we see that when the scale factors are all equal, the non-metricity scalar has a maximum, whereas at the Big Bang/Big Crunch when we have the maximum anisotropy, the non-metricity scalar has a(n infinite) minimum. Similar behaviors might be found in the other figures. While these results suggest the presence of a correlation between non-metricity and anisotropy, we also found a counterexample in Fig. \ref{fig:BI_example2_2}. This exception indicates that non-metricity and anisotropy do not relate straightforwardly. Unfortunately, within the framework of this work, it was not possible to derive a formal relationship between these two quantities. Future works will aim at the clarification of such relation with more sophisticated tools.

We then moved on to the study of spatially flat FLRW universes. In such a framework, we found different solutions: some of them represent  Big Crunch models (Sec. \ref{subsec:FLRW_Example_3}) or oscillating models (Sec. \ref{subsec:FLRW_Example_2}) \cite{Ashtekar_2011, Brandenberger_2017}, where the non-metricity leads the Universe to contract. 
The solution discussed in Sec. \ref{subsec:FLRW_Example_1_2} shows that in the presence of non-metricity, the scale factor can present all the principal phases of the Universe history (inflation, decelerated expansion, and dark era). In this case, it turns out that the non-metricity terms can drive all three phases. 

We conclude remarking that the relation between non-metricity, anisotropy, and expansion of the Universe that we have seen above resembles the one that ties the Ricci scalar to those quantities in metric $f(R)$ gravity. In this respect, therefore, in spite of the reduced degrees of freedom, $f(\mathcal{Q})$ gravity appears to present the same potential of $f(R)$ gravity to generate nontrivial cosmologies and to reproduce cosmic acceleration, at the same time avoiding the issues connected with higher order field equations.

\bibliographystyle{apsrev4-2}
\bibliography{main}

\begin{thebibliography}{41}%
\makeatletter
\providecommand \@ifxundefined [1]{%
 \@ifx{#1\undefined}
}%
\providecommand \@ifnum [1]{%
 \ifnum #1\expandafter \@firstoftwo
 \else \expandafter \@secondoftwo
 \fi
}%
\providecommand \@ifx [1]{%
 \ifx #1\expandafter \@firstoftwo
 \else \expandafter \@secondoftwo
 \fi
}%
\providecommand \natexlab [1]{#1}%
\providecommand \enquote  [1]{``#1''}%
\providecommand \bibnamefont  [1]{#1}%
\providecommand \bibfnamefont [1]{#1}%
\providecommand \citenamefont [1]{#1}%
\providecommand \href@noop [0]{\@secondoftwo}%
\providecommand \href [0]{\begingroup \@sanitize@url \@href}%
\providecommand \@href[1]{\@@startlink{#1}\@@href}%
\providecommand \@@href[1]{\endgroup#1\@@endlink}%
\providecommand \@sanitize@url [0]{\catcode `\\12\catcode `\$12\catcode
  `\&12\catcode `\#12\catcode `\^12\catcode `\_12\catcode `\%12\relax}%
\providecommand \@@startlink[1]{}%
\providecommand \@@endlink[0]{}%
\providecommand \url  [0]{\begingroup\@sanitize@url \@url }%
\providecommand \@url [1]{\endgroup\@href {#1}{\urlprefix }}%
\providecommand \urlprefix  [0]{URL }%
\providecommand \Eprint [0]{\href }%
\providecommand \doibase [0]{https://doi.org/}%
\providecommand \selectlanguage [0]{\@gobble}%
\providecommand \bibinfo  [0]{\@secondoftwo}%
\providecommand \bibfield  [0]{\@secondoftwo}%
\providecommand \translation [1]{[#1]}%
\providecommand \BibitemOpen [0]{}%
\providecommand \bibitemStop [0]{}%
\providecommand \bibitemNoStop [0]{.\EOS\space}%
\providecommand \EOS [0]{\spacefactor3000\relax}%
\providecommand \BibitemShut  [1]{\csname bibitem#1\endcsname}%
\let\auto@bib@innerbib\@empty
\bibitem [{\citenamefont {Wald}(1984)}]{Wald:1984rg}%
  \BibitemOpen
  \bibfield  {author} {\bibinfo {author} {\bibfnamefont {R.~M.}\ \bibnamefont
  {Wald}},\ }\href {https://doi.org/10.7208/chicago/9780226870373.001.0001}
  {\emph {\bibinfo {title} {{General Relativity}}}}\ (\bibinfo  {publisher}
  {Chicago Univ. Pr.},\ \bibinfo {address} {Chicago, USA},\ \bibinfo {year}
  {1984})\BibitemShut {NoStop}%
\bibitem [{\citenamefont {Weinberg}\ and\ \citenamefont
  {Steven}(1972)}]{weinberg1972gravitation}%
  \BibitemOpen
  \bibfield  {author} {\bibinfo {author} {\bibfnamefont {S.}~\bibnamefont
  {Weinberg}}\ and\ \bibinfo {author} {\bibfnamefont {W.}~\bibnamefont
  {Steven}},\ }\href {https://books.google.it/books?id=XLbvAAAAMAAJ} {\emph
  {\bibinfo {title} {Gravitation and Cosmology: Principles and Applications of
  the General Theory of Relativity}}}\ (\bibinfo  {publisher} {Wiley},\
  \bibinfo {year} {1972})\BibitemShut {NoStop}%
\bibitem [{\citenamefont {Carroll}(2001)}]{Carroll_2001}%
  \BibitemOpen
  \bibfield  {author} {\bibinfo {author} {\bibfnamefont {S.~M.}\ \bibnamefont
  {Carroll}},\ }\href {https://doi.org/10.12942/lrr-2001-1} {\bibfield
  {journal} {\bibinfo  {journal} {Living Rev. Rel.}\ }\textbf {\bibinfo
  {volume} {4}},\ \bibinfo {pages} {1} (\bibinfo {year} {2001})}\BibitemShut
  {NoStop}%
\bibitem [{\citenamefont {Peebles}\ and\ \citenamefont
  {Ratra}(2003)}]{Peebles_2003}%
  \BibitemOpen
  \bibfield  {author} {\bibinfo {author} {\bibfnamefont {P.~J.~E.}\
  \bibnamefont {Peebles}}\ and\ \bibinfo {author} {\bibfnamefont
  {B.}~\bibnamefont {Ratra}},\ }\href
  {https://doi.org/10.1103/RevModPhys.75.559} {\bibfield  {journal} {\bibinfo
  {journal} {Rev. Mod. Phys.}\ }\textbf {\bibinfo {volume} {75}},\ \bibinfo
  {pages} {559} (\bibinfo {year} {2003})}\BibitemShut {NoStop}%
\bibitem [{\citenamefont {Bamba}\ \emph {et~al.}(2012)\citenamefont {Bamba},
  \citenamefont {Capozziello}, \citenamefont {Nojiri},\ and\ \citenamefont
  {Odintsov}}]{Bamba:2012cp}%
  \BibitemOpen
  \bibfield  {author} {\bibinfo {author} {\bibfnamefont {K.}~\bibnamefont
  {Bamba}}, \bibinfo {author} {\bibfnamefont {S.}~\bibnamefont {Capozziello}},
  \bibinfo {author} {\bibfnamefont {S.}~\bibnamefont {Nojiri}},\ and\ \bibinfo
  {author} {\bibfnamefont {S.~D.}\ \bibnamefont {Odintsov}},\ }\href
  {https://doi.org/10.1007/s10509-012-1181-8} {\bibfield  {journal} {\bibinfo
  {journal} {Astrophys. Space Sci.}\ }\textbf {\bibinfo {volume} {342}},\
  \bibinfo {pages} {155} (\bibinfo {year} {2012})}\BibitemShut {NoStop}%
\bibitem [{\citenamefont {Weinberg}(1989)}]{RevModPhys.61.1}%
  \BibitemOpen
  \bibfield  {author} {\bibinfo {author} {\bibfnamefont {S.}~\bibnamefont
  {Weinberg}},\ }\href {https://doi.org/10.1103/RevModPhys.61.1} {\bibfield
  {journal} {\bibinfo  {journal} {Rev. Mod. Phys.}\ }\textbf {\bibinfo {volume}
  {61}},\ \bibinfo {pages} {1} (\bibinfo {year} {1989})}\BibitemShut {NoStop}%
\bibitem [{\citenamefont {Capozziello}\ and\ \citenamefont
  {De~Laurentis}(2011)}]{CAPOZZIELLO2011167}%
  \BibitemOpen
  \bibfield  {author} {\bibinfo {author} {\bibfnamefont {S.}~\bibnamefont
  {Capozziello}}\ and\ \bibinfo {author} {\bibfnamefont {M.}~\bibnamefont
  {De~Laurentis}},\ }\href {https://doi.org/10.1016/j.physrep.2011.09.003}
  {\bibfield  {journal} {\bibinfo  {journal} {Phys. Rept.}\ }\textbf {\bibinfo
  {volume} {509}},\ \bibinfo {pages} {167} (\bibinfo {year}
  {2011})}\BibitemShut {NoStop}%
\bibitem [{\citenamefont {Sotiriou}\ and\ \citenamefont
  {Faraoni}(2010)}]{Sotiriou_2010}%
  \BibitemOpen
  \bibfield  {author} {\bibinfo {author} {\bibfnamefont {T.~P.}\ \bibnamefont
  {Sotiriou}}\ and\ \bibinfo {author} {\bibfnamefont {V.}~\bibnamefont
  {Faraoni}},\ }\href {https://doi.org/10.1103/RevModPhys.82.451} {\bibfield
  {journal} {\bibinfo  {journal} {Rev. Mod. Phys.}\ }\textbf {\bibinfo {volume}
  {82}},\ \bibinfo {pages} {451} (\bibinfo {year} {2010})}\BibitemShut
  {NoStop}%
\bibitem [{\citenamefont {Clifton}\ \emph {et~al.}(2012)\citenamefont
  {Clifton}, \citenamefont {Ferreira}, \citenamefont {Padilla},\ and\
  \citenamefont {Skordis}}]{Clifton_2012}%
  \BibitemOpen
  \bibfield  {author} {\bibinfo {author} {\bibfnamefont {T.}~\bibnamefont
  {Clifton}}, \bibinfo {author} {\bibfnamefont {P.~G.}\ \bibnamefont
  {Ferreira}}, \bibinfo {author} {\bibfnamefont {A.}~\bibnamefont {Padilla}},\
  and\ \bibinfo {author} {\bibfnamefont {C.}~\bibnamefont {Skordis}},\ }\href
  {https://doi.org/10.1016/j.physrep.2012.01.001} {\bibfield  {journal}
  {\bibinfo  {journal} {Phys. Rept.}\ }\textbf {\bibinfo {volume} {513}},\
  \bibinfo {pages} {1} (\bibinfo {year} {2012})}\BibitemShut {NoStop}%
\bibitem [{\citenamefont {Jim\'enez}\ \emph {et~al.}(2019)\citenamefont
  {Jim\'enez}, \citenamefont {Heisenberg},\ and\ \citenamefont
  {Koivisto}}]{BeltranJimenez:2019tjy}%
  \BibitemOpen
  \bibfield  {author} {\bibinfo {author} {\bibfnamefont {J.~B.}\ \bibnamefont
  {Jim\'enez}}, \bibinfo {author} {\bibfnamefont {L.}~\bibnamefont
  {Heisenberg}},\ and\ \bibinfo {author} {\bibfnamefont {T.~S.}\ \bibnamefont
  {Koivisto}},\ }\href {https://doi.org/10.3390/universe5070173} {\bibfield
  {journal} {\bibinfo  {journal} {Universe}\ }\textbf {\bibinfo {volume} {5}},\
  \bibinfo {pages} {173} (\bibinfo {year} {2019})}\BibitemShut {NoStop}%
\bibitem [{\citenamefont {Kr{\v{s}}{\v{s}}\'{a}k}\ \emph
  {et~al.}(2019)\citenamefont {Kr{\v{s}}{\v{s}}\'{a}k}, \citenamefont {van~den
  Hoogen}, \citenamefont {Pereira}, \citenamefont {B{\"o}hmer},\ and\
  \citenamefont {Coley}}]{Kr_k_2019}%
  \BibitemOpen
  \bibfield  {author} {\bibinfo {author} {\bibfnamefont {M.}~\bibnamefont
  {Kr{\v{s}}{\v{s}}\'{a}k}}, \bibinfo {author} {\bibfnamefont {R.~J.}\
  \bibnamefont {van~den Hoogen}}, \bibinfo {author} {\bibfnamefont {J.~G.}\
  \bibnamefont {Pereira}}, \bibinfo {author} {\bibfnamefont {C.~G.}\
  \bibnamefont {B{\"o}hmer}},\ and\ \bibinfo {author} {\bibfnamefont {A.~A.}\
  \bibnamefont {Coley}},\ }\href {https://doi.org/10.1088/1361-6382/ab2e1f}
  {\bibfield  {journal} {\bibinfo  {journal} {Class. Quant. Grav.}\ }\textbf
  {\bibinfo {volume} {36}},\ \bibinfo {pages} {183001} (\bibinfo {year}
  {2019})}\BibitemShut {NoStop}%
\bibitem [{\citenamefont {Nester}\ and\ \citenamefont
  {Yo}(1999)}]{Nester:1998mp}%
  \BibitemOpen
  \bibfield  {author} {\bibinfo {author} {\bibfnamefont {J.~M.}\ \bibnamefont
  {Nester}}\ and\ \bibinfo {author} {\bibfnamefont {H.-J.}\ \bibnamefont
  {Yo}},\ }\href@noop {} {\bibfield  {journal} {\bibinfo  {journal} {Chin. J.
  Phys.}\ }\textbf {\bibinfo {volume} {37}},\ \bibinfo {pages} {113} (\bibinfo
  {year} {1999})},\ \Eprint {https://arxiv.org/abs/gr-qc/9809049}
  {arXiv:gr-qc/9809049 [gr-qc]} \BibitemShut {NoStop}%
\bibitem [{\citenamefont {Adak}\ \emph {et~al.}(2013)\citenamefont {Adak},
  \citenamefont {Sert}, \citenamefont {Kalay},\ and\ \citenamefont
  {Sari}}]{Adak:2008gd}%
  \BibitemOpen
  \bibfield  {author} {\bibinfo {author} {\bibfnamefont {M.}~\bibnamefont
  {Adak}}, \bibinfo {author} {\bibfnamefont {{\"O}.}~\bibnamefont {Sert}},
  \bibinfo {author} {\bibfnamefont {M.}~\bibnamefont {Kalay}},\ and\ \bibinfo
  {author} {\bibfnamefont {M.}~\bibnamefont {Sari}},\ }\href
  {https://doi.org/10.1142/S0217751X13501674} {\bibfield  {journal} {\bibinfo
  {journal} {Int. J. Mod. Phys. A}\ }\textbf {\bibinfo {volume} {28}},\
  \bibinfo {pages} {1350167} (\bibinfo {year} {2013})}\BibitemShut {NoStop}%
\bibitem [{\citenamefont {Adak}\ \emph {et~al.}(2006)\citenamefont {Adak},
  \citenamefont {Kalay},\ and\ \citenamefont {Sert}}]{Adak:2005cd}%
  \BibitemOpen
  \bibfield  {author} {\bibinfo {author} {\bibfnamefont {M.}~\bibnamefont
  {Adak}}, \bibinfo {author} {\bibfnamefont {M.}~\bibnamefont {Kalay}},\ and\
  \bibinfo {author} {\bibfnamefont {O.}~\bibnamefont {Sert}},\ }\href
  {https://doi.org/10.1142/S0218271806008474} {\bibfield  {journal} {\bibinfo
  {journal} {Int. J. Mod. Phys.}\ }\textbf {\bibinfo {volume} {D15}},\ \bibinfo
  {pages} {619} (\bibinfo {year} {2006})}\BibitemShut {NoStop}%
\bibitem [{\citenamefont {Bombacigno}\ \emph {et~al.}(2021)\citenamefont
  {Bombacigno}, \citenamefont {Boudet}, \citenamefont {Olmo},\ and\
  \citenamefont {Montani}}]{Bombacigno:2021bpk}%
  \BibitemOpen
  \bibfield  {author} {\bibinfo {author} {\bibfnamefont {F.}~\bibnamefont
  {Bombacigno}}, \bibinfo {author} {\bibfnamefont {S.}~\bibnamefont {Boudet}},
  \bibinfo {author} {\bibfnamefont {G.~J.}\ \bibnamefont {Olmo}},\ and\
  \bibinfo {author} {\bibfnamefont {G.}~\bibnamefont {Montani}},\ }\href
  {https://doi.org/10.1103/PhysRevD.103.124031} {\bibfield  {journal} {\bibinfo
   {journal} {Phys. Rev. D}\ }\textbf {\bibinfo {volume} {103}},\ \bibinfo
  {pages} {124031} (\bibinfo {year} {2021})}\BibitemShut {NoStop}%
\bibitem [{\citenamefont {Conroy}\ and\ \citenamefont
  {Koivisto}(2018)}]{Conroy:2017yln}%
  \BibitemOpen
  \bibfield  {author} {\bibinfo {author} {\bibfnamefont {A.}~\bibnamefont
  {Conroy}}\ and\ \bibinfo {author} {\bibfnamefont {T.}~\bibnamefont
  {Koivisto}},\ }\href {https://doi.org/10.1140/epjc/s10052-018-6410-z}
  {\bibfield  {journal} {\bibinfo  {journal} {Eur. Phys. J. C}\ }\textbf
  {\bibinfo {volume} {78}},\ \bibinfo {pages} {923} (\bibinfo {year}
  {2018})}\BibitemShut {NoStop}%
\bibitem [{\citenamefont {Cai}\ \emph {et~al.}(2016)\citenamefont {Cai},
  \citenamefont {Capozziello}, \citenamefont {De~Laurentis},\ and\
  \citenamefont {Saridakis}}]{Cai:2015emx}%
  \BibitemOpen
  \bibfield  {author} {\bibinfo {author} {\bibfnamefont {Y.-F.}\ \bibnamefont
  {Cai}}, \bibinfo {author} {\bibfnamefont {S.}~\bibnamefont {Capozziello}},
  \bibinfo {author} {\bibfnamefont {M.}~\bibnamefont {De~Laurentis}},\ and\
  \bibinfo {author} {\bibfnamefont {E.~N.}\ \bibnamefont {Saridakis}},\ }\href
  {https://doi.org/10.1088/0034-4885/79/10/106901} {\bibfield  {journal}
  {\bibinfo  {journal} {Rept. Prog. Phys.}\ }\textbf {\bibinfo {volume} {79}},\
  \bibinfo {pages} {106901} (\bibinfo {year} {2016})}\BibitemShut {NoStop}%
\bibitem [{\citenamefont {Jim\'enez}\ \emph {et~al.}(2018)\citenamefont
  {Jim\'enez}, \citenamefont {Heisenberg},\ and\ \citenamefont
  {Koivisto}}]{BeltranJimenez:2017tkd}%
  \BibitemOpen
  \bibfield  {author} {\bibinfo {author} {\bibfnamefont {J.~B.}\ \bibnamefont
  {Jim\'enez}}, \bibinfo {author} {\bibfnamefont {L.}~\bibnamefont
  {Heisenberg}},\ and\ \bibinfo {author} {\bibfnamefont {T.~S.}\ \bibnamefont
  {Koivisto}},\ }\href {https://doi.org/10.1103/PhysRevD.98.044048} {\bibfield
  {journal} {\bibinfo  {journal} {Phys. Rev. D}\ }\textbf {\bibinfo {volume}
  {98}},\ \bibinfo {pages} {044048} (\bibinfo {year} {2018})}\BibitemShut
  {NoStop}%
\bibitem [{\citenamefont {Motohashi}\ and\ \citenamefont
  {Suyama}(2015)}]{Motohashi_2015}%
  \BibitemOpen
  \bibfield  {author} {\bibinfo {author} {\bibfnamefont {H.}~\bibnamefont
  {Motohashi}}\ and\ \bibinfo {author} {\bibfnamefont {T.}~\bibnamefont
  {Suyama}},\ }\href {https://doi.org/10.1103/PhysRevD.91.085009} {\bibfield
  {journal} {\bibinfo  {journal} {Phys. Rev. D}\ }\textbf {\bibinfo {volume}
  {91}},\ \bibinfo {pages} {085009} (\bibinfo {year} {2015})}\BibitemShut
  {NoStop}%
\bibitem [{\citenamefont {Woodard}(2015)}]{Woodard:2015zca}%
  \BibitemOpen
  \bibfield  {author} {\bibinfo {author} {\bibfnamefont {R.~P.}\ \bibnamefont
  {Woodard}},\ }\href {https://doi.org/10.4249/scholarpedia.32243} {\bibfield
  {journal} {\bibinfo  {journal} {Scholarpedia}\ }\textbf {\bibinfo {volume}
  {10}},\ \bibinfo {pages} {32243} (\bibinfo {year} {2015})},\ \Eprint
  {https://arxiv.org/abs/1506.02210} {arXiv:1506.02210 [hep-th]} \BibitemShut
  {NoStop}%
\bibitem [{\citenamefont {Harko}\ \emph {et~al.}(2018)\citenamefont {Harko},
  \citenamefont {Koivisto}, \citenamefont {Lobo}, \citenamefont {Olmo},\ and\
  \citenamefont {Rubiera-Garcia}}]{Harko_2018}%
  \BibitemOpen
  \bibfield  {author} {\bibinfo {author} {\bibfnamefont {T.}~\bibnamefont
  {Harko}}, \bibinfo {author} {\bibfnamefont {T.~S.}\ \bibnamefont {Koivisto}},
  \bibinfo {author} {\bibfnamefont {F.~S.~N.}\ \bibnamefont {Lobo}}, \bibinfo
  {author} {\bibfnamefont {G.~J.}\ \bibnamefont {Olmo}},\ and\ \bibinfo
  {author} {\bibfnamefont {D.}~\bibnamefont {Rubiera-Garcia}},\ }\href
  {https://doi.org/10.1103/PhysRevD.98.084043} {\bibfield  {journal} {\bibinfo
  {journal} {Phys. Rev. D}\ }\textbf {\bibinfo {volume} {98}},\ \bibinfo
  {pages} {084043} (\bibinfo {year} {2018})}\BibitemShut {NoStop}%
\bibitem [{\citenamefont {Beltr\'an~Jim\'enez}\ \emph
  {et~al.}(2020)\citenamefont {Beltr\'an~Jim\'enez}, \citenamefont
  {Heisenberg}, \citenamefont {Koivisto},\ and\ \citenamefont
  {Pekar}}]{Jim_nez_2020}%
  \BibitemOpen
  \bibfield  {author} {\bibinfo {author} {\bibfnamefont {J.}~\bibnamefont
  {Beltr\'an~Jim\'enez}}, \bibinfo {author} {\bibfnamefont {L.}~\bibnamefont
  {Heisenberg}}, \bibinfo {author} {\bibfnamefont {T.~S.}\ \bibnamefont
  {Koivisto}},\ and\ \bibinfo {author} {\bibfnamefont {S.}~\bibnamefont
  {Pekar}},\ }\href {https://doi.org/10.1103/PhysRevD.101.103507} {\bibfield
  {journal} {\bibinfo  {journal} {Phys. Rev. D}\ }\textbf {\bibinfo {volume}
  {101}},\ \bibinfo {pages} {103507} (\bibinfo {year} {2020})}\BibitemShut
  {NoStop}%
\bibitem [{\citenamefont {Mandal}\ \emph {et~al.}(2020)\citenamefont {Mandal},
  \citenamefont {Sahoo},\ and\ \citenamefont {Santos}}]{Mandal_2020}%
  \BibitemOpen
  \bibfield  {author} {\bibinfo {author} {\bibfnamefont {S.}~\bibnamefont
  {Mandal}}, \bibinfo {author} {\bibfnamefont {P.~K.}\ \bibnamefont {Sahoo}},\
  and\ \bibinfo {author} {\bibfnamefont {J.~R.~L.}\ \bibnamefont {Santos}},\
  }\href {https://doi.org/10.1103/PhysRevD.102.024057} {\bibfield  {journal}
  {\bibinfo  {journal} {Phys. Rev. D}\ }\textbf {\bibinfo {volume} {102}},\
  \bibinfo {pages} {024057} (\bibinfo {year} {2020})}\BibitemShut {NoStop}%
\bibitem [{\citenamefont {Barros}\ \emph {et~al.}(2020)\citenamefont {Barros},
  \citenamefont {Barreiro}, \citenamefont {Koivisto},\ and\ \citenamefont
  {Nunes}}]{Barros_2020}%
  \BibitemOpen
  \bibfield  {author} {\bibinfo {author} {\bibfnamefont {B.~J.}\ \bibnamefont
  {Barros}}, \bibinfo {author} {\bibfnamefont {T.}~\bibnamefont {Barreiro}},
  \bibinfo {author} {\bibfnamefont {T.}~\bibnamefont {Koivisto}},\ and\
  \bibinfo {author} {\bibfnamefont {N.~J.}\ \bibnamefont {Nunes}},\ }\href
  {https://doi.org/10.1016/j.dark.2020.100616} {\bibfield  {journal} {\bibinfo
  {journal} {Phys. Dark Univ.}\ }\textbf {\bibinfo {volume} {30}},\ \bibinfo
  {pages} {100616} (\bibinfo {year} {2020})}\BibitemShut {NoStop}%
\bibitem [{\citenamefont {Lazkoz}\ \emph {et~al.}(2019)\citenamefont {Lazkoz},
  \citenamefont {Lobo}, \citenamefont {Ortiz-Ba\~nos},\ and\ \citenamefont
  {Salzano}}]{Lazkoz_2019}%
  \BibitemOpen
  \bibfield  {author} {\bibinfo {author} {\bibfnamefont {R.}~\bibnamefont
  {Lazkoz}}, \bibinfo {author} {\bibfnamefont {F.~S.~N.}\ \bibnamefont {Lobo}},
  \bibinfo {author} {\bibfnamefont {M.}~\bibnamefont {Ortiz-Ba\~nos}},\ and\
  \bibinfo {author} {\bibfnamefont {V.}~\bibnamefont {Salzano}},\ }\href
  {https://doi.org/10.1103/PhysRevD.100.104027} {\bibfield  {journal} {\bibinfo
   {journal} {Phys. Rev. D}\ }\textbf {\bibinfo {volume} {100}},\ \bibinfo
  {pages} {104027} (\bibinfo {year} {2019})}\BibitemShut {NoStop}%
\bibitem [{\citenamefont {Soudi}\ \emph {et~al.}(2019)\citenamefont {Soudi},
  \citenamefont {Farrugia}, \citenamefont {Gakis}, \citenamefont {Levi~Said},\
  and\ \citenamefont {Saridakis}}]{Soudi:2018dhv}%
  \BibitemOpen
  \bibfield  {author} {\bibinfo {author} {\bibfnamefont {I.}~\bibnamefont
  {Soudi}}, \bibinfo {author} {\bibfnamefont {G.}~\bibnamefont {Farrugia}},
  \bibinfo {author} {\bibfnamefont {V.}~\bibnamefont {Gakis}}, \bibinfo
  {author} {\bibfnamefont {J.}~\bibnamefont {Levi~Said}},\ and\ \bibinfo
  {author} {\bibfnamefont {E.~N.}\ \bibnamefont {Saridakis}},\ }\href
  {https://doi.org/10.1103/PhysRevD.100.044008} {\bibfield  {journal} {\bibinfo
   {journal} {Phys. Rev. D}\ }\textbf {\bibinfo {volume} {100}},\ \bibinfo
  {pages} {044008} (\bibinfo {year} {2019})}\BibitemShut {NoStop}%
\bibitem [{\citenamefont {Frusciante}(2021)}]{Frusciante_2021}%
  \BibitemOpen
  \bibfield  {author} {\bibinfo {author} {\bibfnamefont {N.}~\bibnamefont
  {Frusciante}},\ }\href {https://doi.org/10.1103/PhysRevD.103.044021}
  {\bibfield  {journal} {\bibinfo  {journal} {Phys. Rev. D}\ }\textbf {\bibinfo
  {volume} {103}},\ \bibinfo {pages} {044021} (\bibinfo {year}
  {2021})}\BibitemShut {NoStop}%
\bibitem [{\citenamefont {Hassan}\ \emph {et~al.}(2021)\citenamefont {Hassan},
  \citenamefont {Mandal},\ and\ \citenamefont {Sahoo}}]{Hassan_2021}%
  \BibitemOpen
  \bibfield  {author} {\bibinfo {author} {\bibfnamefont {Z.}~\bibnamefont
  {Hassan}}, \bibinfo {author} {\bibfnamefont {S.}~\bibnamefont {Mandal}},\
  and\ \bibinfo {author} {\bibfnamefont {P.~K.}\ \bibnamefont {Sahoo}},\ }\href
  {https://doi.org/10.1002/prop.202100023} {\bibfield  {journal} {\bibinfo
  {journal} {Fortsch. Phys.}\ }\textbf {\bibinfo {volume} {69}},\ \bibinfo
  {pages} {2100023} (\bibinfo {year} {2021})}\BibitemShut {NoStop}%
\bibitem [{\citenamefont {Bajardi}\ \emph {et~al.}(2020)\citenamefont
  {Bajardi}, \citenamefont {Vernieri},\ and\ \citenamefont
  {Capozziello}}]{Bajardi_2020}%
  \BibitemOpen
  \bibfield  {author} {\bibinfo {author} {\bibfnamefont {F.}~\bibnamefont
  {Bajardi}}, \bibinfo {author} {\bibfnamefont {D.}~\bibnamefont {Vernieri}},\
  and\ \bibinfo {author} {\bibfnamefont {S.}~\bibnamefont {Capozziello}},\
  }\href {https://doi.org/10.1140/epjp/s13360-020-00918-3} {\bibfield
  {journal} {\bibinfo  {journal} {Eur. Phys. J. Plus}\ }\textbf {\bibinfo
  {volume} {135}},\ \bibinfo {pages} {912} (\bibinfo {year}
  {2020})}\BibitemShut {NoStop}%
\bibitem [{\citenamefont {Ellis}\ and\ \citenamefont
  {Madsen}(1991)}]{Ellis_1991}%
  \BibitemOpen
  \bibfield  {author} {\bibinfo {author} {\bibfnamefont {G.~F.~R.}\
  \bibnamefont {Ellis}}\ and\ \bibinfo {author} {\bibfnamefont {M.~S.}\
  \bibnamefont {Madsen}},\ }\href {https://doi.org/10.1088/0264-9381/8/4/012}
  {\bibfield  {journal} {\bibinfo  {journal} {Class. Quant. Grav.}\ }\textbf
  {\bibinfo {volume} {8}},\ \bibinfo {pages} {667} (\bibinfo {year}
  {1991})}\BibitemShut {NoStop}%
\bibitem [{\citenamefont {Carloni}\ \emph {et~al.}(2012)\citenamefont
  {Carloni}, \citenamefont {Goswami},\ and\ \citenamefont
  {Dunsby}}]{Carloni_2012}%
  \BibitemOpen
  \bibfield  {author} {\bibinfo {author} {\bibfnamefont {S.}~\bibnamefont
  {Carloni}}, \bibinfo {author} {\bibfnamefont {R.}~\bibnamefont {Goswami}},\
  and\ \bibinfo {author} {\bibfnamefont {P.~K.~S.}\ \bibnamefont {Dunsby}},\
  }\href {https://doi.org/10.1088/0264-9381/29/13/135012} {\bibfield  {journal}
  {\bibinfo  {journal} {Class. Quant. Grav.}\ }\textbf {\bibinfo {volume}
  {29}},\ \bibinfo {pages} {135012} (\bibinfo {year} {2012})}\BibitemShut
  {NoStop}%
\bibitem [{\citenamefont {Vignolo}\ \emph {et~al.}(2013)\citenamefont
  {Vignolo}, \citenamefont {Carloni},\ and\ \citenamefont
  {Vietri}}]{Vignolo_2013}%
  \BibitemOpen
  \bibfield  {author} {\bibinfo {author} {\bibfnamefont {S.}~\bibnamefont
  {Vignolo}}, \bibinfo {author} {\bibfnamefont {S.}~\bibnamefont {Carloni}},\
  and\ \bibinfo {author} {\bibfnamefont {F.}~\bibnamefont {Vietri}},\ }\href
  {https://doi.org/10.1103/PhysRevD.88.023006} {\bibfield  {journal} {\bibinfo
  {journal} {Phys. Rev. D}\ }\textbf {\bibinfo {volume} {88}},\ \bibinfo
  {pages} {023006} (\bibinfo {year} {2013})}\BibitemShut {NoStop}%
\bibitem [{\citenamefont {Carloni}\ and\ \citenamefont
  {Vernieri}(2018{\natexlab{a}})}]{Carloni:2017rpu}%
  \BibitemOpen
  \bibfield  {author} {\bibinfo {author} {\bibfnamefont {S.}~\bibnamefont
  {Carloni}}\ and\ \bibinfo {author} {\bibfnamefont {D.}~\bibnamefont
  {Vernieri}},\ }\href {https://doi.org/10.1103/PhysRevD.97.124056} {\bibfield
  {journal} {\bibinfo  {journal} {Phys. Rev. D}\ }\textbf {\bibinfo {volume}
  {97}},\ \bibinfo {pages} {124056} (\bibinfo {year}
  {2018}{\natexlab{a}})}\BibitemShut {NoStop}%
\bibitem [{\citenamefont {Carloni}\ and\ \citenamefont
  {Vernieri}(2018{\natexlab{b}})}]{Carloni:2017bck}%
  \BibitemOpen
  \bibfield  {author} {\bibinfo {author} {\bibfnamefont {S.}~\bibnamefont
  {Carloni}}\ and\ \bibinfo {author} {\bibfnamefont {D.}~\bibnamefont
  {Vernieri}},\ }\href {https://doi.org/10.1103/PhysRevD.97.124057} {\bibfield
  {journal} {\bibinfo  {journal} {Phys. Rev. D}\ }\textbf {\bibinfo {volume}
  {97}},\ \bibinfo {pages} {124057} (\bibinfo {year}
  {2018}{\natexlab{b}})}\BibitemShut {NoStop}%
\bibitem [{\citenamefont {Carloni}(2014)}]{Carloni:2014rba}%
  \BibitemOpen
  \bibfield  {author} {\bibinfo {author} {\bibfnamefont {S.}~\bibnamefont
  {Carloni}},\ }\href {https://doi.org/10.1103/PhysRevD.90.044023} {\bibfield
  {journal} {\bibinfo  {journal} {Phys. Rev. D}\ }\textbf {\bibinfo {volume}
  {90}},\ \bibinfo {pages} {044023} (\bibinfo {year} {2014})}\BibitemShut
  {NoStop}%
\bibitem [{\citenamefont {{Naidu}}\ \emph {et~al.}(2021)\citenamefont
  {{Naidu}}, \citenamefont {{Carloni}},\ and\ \citenamefont
  {{Dunsby}}}]{2021arXiv210205693N}%
  \BibitemOpen
  \bibfield  {author} {\bibinfo {author} {\bibfnamefont {N.~F.}\ \bibnamefont
  {{Naidu}}}, \bibinfo {author} {\bibfnamefont {S.}~\bibnamefont {{Carloni}}},\
  and\ \bibinfo {author} {\bibfnamefont {P.}~\bibnamefont {{Dunsby}}},\
  }\href@noop {} {\bibfield  {journal} {\bibinfo  {journal} {arXiv e-prints}\ }
  (\bibinfo {year} {2021})},\ \Eprint {https://arxiv.org/abs/2102.05693}
  {arXiv:2102.05693 [gr-qc]} \BibitemShut {NoStop}%
\bibitem [{\citenamefont {Ellis}\ and\ \citenamefont
  {MacCallum}(1969)}]{Ellis1969}%
  \BibitemOpen
  \bibfield  {author} {\bibinfo {author} {\bibfnamefont {G.~F.~R.}\
  \bibnamefont {Ellis}}\ and\ \bibinfo {author} {\bibfnamefont {M.~A.~H.}\
  \bibnamefont {MacCallum}},\ }\href {https://doi.org/10.1007/BF01645908}
  {\bibfield  {journal} {\bibinfo  {journal} {Commun. Math. Phys.}\ }\textbf
  {\bibinfo {volume} {12}},\ \bibinfo {pages} {108} (\bibinfo {year}
  {1969})}\BibitemShut {NoStop}%
\bibitem [{\citenamefont {Ellis}(2006)}]{Ellis2006}%
  \BibitemOpen
  \bibfield  {author} {\bibinfo {author} {\bibfnamefont {G.~F.~R.}\
  \bibnamefont {Ellis}},\ }\href {https://doi.org/10.1007/s10714-006-0283-4}
  {\bibfield  {journal} {\bibinfo  {journal} {Gen. Rel. Grav.}\ }\textbf
  {\bibinfo {volume} {38}},\ \bibinfo {pages} {1003} (\bibinfo {year}
  {2006})}\BibitemShut {NoStop}%
\bibitem [{\citenamefont {Stephani}(2004)}]{stephani2004relativity}%
  \BibitemOpen
  \bibfield  {author} {\bibinfo {author} {\bibfnamefont {H.}~\bibnamefont
  {Stephani}},\ }\href {https://books.google.it/books?id=WAW-4nd-OeIC} {\emph
  {\bibinfo {title} {Relativity: An Introduction to Special and General
  Relativity}}}\ (\bibinfo  {publisher} {Cambridge University Press},\ \bibinfo
  {year} {2004})\BibitemShut {NoStop}%
\bibitem [{\citenamefont {Ashtekar}\ and\ \citenamefont
  {Singh}(2011)}]{Ashtekar_2011}%
  \BibitemOpen
  \bibfield  {author} {\bibinfo {author} {\bibfnamefont {A.}~\bibnamefont
  {Ashtekar}}\ and\ \bibinfo {author} {\bibfnamefont {P.}~\bibnamefont
  {Singh}},\ }\href {https://doi.org/10.1088/0264-9381/28/21/213001} {\bibfield
   {journal} {\bibinfo  {journal} {Class. Quant. Grav.}\ }\textbf {\bibinfo
  {volume} {28}},\ \bibinfo {pages} {213001} (\bibinfo {year}
  {2011})}\BibitemShut {NoStop}%
\bibitem [{\citenamefont {Brandenberger}\ and\ \citenamefont
  {Peter}(2017)}]{Brandenberger_2017}%
  \BibitemOpen
  \bibfield  {author} {\bibinfo {author} {\bibfnamefont {R.}~\bibnamefont
  {Brandenberger}}\ and\ \bibinfo {author} {\bibfnamefont {P.}~\bibnamefont
  {Peter}},\ }\href {https://doi.org/10.1007/s10701-016-0057-0} {\bibfield
  {journal} {\bibinfo  {journal} {Found. Phys.}\ }\textbf {\bibinfo {volume}
  {47}},\ \bibinfo {pages} {797} (\bibinfo {year} {2017})}\BibitemShut
  {NoStop}%
\end{thebibliography}%

\end{document}